\documentclass{JFM-FLM_Au}
\usepackage{subcaption}
\usepackage{multirow}
\usepackage{xcolor}
\usepackage{overpic}
\usepackage{enumitem}
\usepackage{hyperref}
\hypersetup{
    colorlinks = true,
    linkcolor  = blue,
    filecolor  = blue,
    urlcolor   = blue,
    citecolor  = blue,
}

\lefttitle{P. Jiang \textit{et al.}}
\righttitle{Pressure fluctuations of flow over a fully appended submarine model}

\title{Experimental investigation of wall-pressure fluctuations on a fully appended submarine model at high Reynolds numbers}

\author{
Peng Jiang\aff{1},
Haoyu Zhang\aff{1},
Yi Dai\aff{1,2},
Tao Peng\aff{1,2},
Bin Xie\aff{1,2}
\and Shijun Liao\aff{1,2}
}

\affiliation{
\aff{1}School of Ocean and Civil Engineering, Shanghai Jiao Tong University, Shanghai, 200240, China
\aff{2}State Key Laboratory of Ocean Engineering, Shanghai Jiao Tong University, Shanghai, 200240, China}

\corresau{Shijun Liao, \email{sjliao@sjtu.edu.cn}}

\begin{document}
\maketitle

\begin{abstract}
This paper \textcolor{black}{addresses a critical gap in hydroacoustics through a systematic wind tunnel investigation of wall-pressure fluctuations on the fully appended DARPA SUBOFF model at operationally relevant Reynolds numbers ranging from $5.6 \times 10^{6}$ to $1.4 \times 10^{7}$.} The experimental campaign encompasses baseline straight-ahead flow, complex maneuvering (yaw and pitch) conditions, and a first-of-its-kind assessment of a novel vortex control baffle (VCB). To ensure benchmark-quality spectral data, rigorous signal processing techniques were applied, specifically Wiener filtering for background noise suppression and dynamic transfer function correction for pinhole sensors. \textcolor{black}{Key findings indicate that while spectral self-similarity holds across Reynolds numbers, the primary finding is the critical role of appendages in noise amplification. Unstable horseshoe vortex dynamics at the sail-hull junction drive localized pressure fluctuations of up to 300\%, establishing this feature as a major coherent noise source. To address this, the study provides the pioneering experimental validation of the VCB. By physically suppressing horseshoe vortex formation at the sail-hull junction, the VCB achieves a global stabilization of the downstream flow, resulting in a significant 35\% reduction in root-mean-square wall-pressure fluctuations at the stern and an approximately 14\% reduction along the parallel mid-body. Furthermore, maneuvering conditions are shown to fundamentally reshape the pressure field, introducing substantial crossflow effects and non-monotonic spectral behaviors. This comprehensive dataset and the demonstrated efficacy of the VCB provide essential physical insights and a critical validation benchmark for the design of next-generation quiet submarines.}
\end{abstract}


\section{Introduction}\label{introduction}
Hydrodynamic stealth, especially for submarines, remains a critical challenge in modern naval engineering \citep{Yu2007shipNoise, Blake2017Mechanics}. 
The total radiated noise signature comprises three distinct categories: machinery noise, propeller noise, and hydrodynamic noise \citep{Yu2007shipNoise}. 
Although advances in vibration isolation and manufacturing have significantly reduced the first two, hydrodynamic noise persists as a difficult acoustic problem. 
This category is theoretically divided into two mechanisms: flow noise, the direct acoustic radiation from turbulent instabilities, and flow-induced noise, the sound re-radiated from structural vibrations excited by unsteady surface pressures \citep{zhang2020review, Jia2022Experimental}. While less intense at low speeds, hydrodynamic noise scales strongly with velocity, typically with the sixth to eighth power ($\sim U^6$--$U^8$), making it the dominant acoustic source during high-speed operations \citep{Yu2007shipNoise, Ciappi2005characteristics}. Crucially, both mechanisms originate from a singular physical source: unsteady wall-pressure fluctuations (WPF) within the turbulent boundary layer (TBL) \citep{Willmarth1975Pressure, Yang_Yang_2022}. Consequently, the accurate characterization of these fluctuations is a foundational requirement for the predictive modeling and acoustic optimization of next-generation quiet submarines \citep{Jiang2024OEHUll, Jiang2025PoFSUBOFF, Zhou2022suboff}.

The theoretical foundation for analyzing wall-pressure fluctuations is well-established for canonical, zero-pressure-gradient (ZPG) boundary layers. Seminal works by \citet{kraichnan1964kolmogorovs} and  \citet{corcos1964structure} established the theoretical framework by analytically linking surface pressure spectra to turbulent velocity field statistics. These theoretical insights were later refined into robust semi-empirical models by \citet{farabee1991spectral} and \citet{Goody2004}, which successfully collapse experimental data for \textcolor{black}{ZPG boundary flows.} However, the hydrodynamic environment of an operational submarine deviates fundamentally from these idealized flat-plate baselines. The flow field is dominated by complex three-dimensional curvature, spatially developing pressure gradients-notably the strong \textcolor{black}{adverse pressure gradient (APG)} at the stern-and, most critically, the highly energetic wakes and junction vortices generated by appendages. Recent investigations have highlighted the breakdown of classical scaling laws in these non-equilibrium regimes. For instance, recent study by \citet{Fan2025ScalingJFM} demonstrated that in the stern region, the classical scaling based on wall shear stress ($\tau_w$) fails to collapse the pressure spectra, suggesting that alternative parameters like the local maximum Reynolds shear stress may be necessary. \textcolor{black}{Consequently, simply extrapolating canonical models is insufficient for predicting the acoustic signature of a fully appended hull. In the absence of universal scaling laws, researchers have increasingly turned to Computational Fluid Dynamics (CFD); however, numerical approaches face significant challenges at operationally relevant conditions. High-fidelity methods, particularly Large Eddy Simulation (LES), have provided valuable insights into flow topology \citep{posa2016suboff, MorseMahesh2023Trippingeffects}. Yet, resolving the full frequency spectrum of wall-pressure fluctuations at Reynolds numbers exceeding $10^7$ remains computationally prohibitive. While wall-modeled LES (WMLES) offer a feasible compromise, their predictive accuracy in regions of massive separation-such as the sail-hull junction-relies heavily on validation based on experimental data to rule out modeling artifacts \citep{Jiang2024OEHUll, Jiang2025PoFSUBOFF}. Therefore, high-fidelity experimental measurements are needed, serving not only to investigate flow physics but also as a rigorous validation benchmark for next-generation numerical models.}

Despite this critical need, a sharp contrast exists between the availability of mean and unsteady flow databases. On the one hand, the mean hydrodynamic field is well-documented, with benchmarks established for \textcolor{black}{mean pressure distributions \citep{Liu1998SUBOFFexp} and mean velocity and Reynolds stress distributions \citep{Jimenez2010JFMwakeflow}.} On the other hand, high-fidelity data for unsteady wall-pressure fluctuations remain surprisingly limited. Existing databases are largely confined to low-to-moderate Reynolds numbers or simplified `bare hull' geometries \citep{Jimenez2010JFMwakeflow, Balantrapu2023APG}, failing to capture the nonlinear interaction between appendage-induced vortices. \textcolor{black}{This lack of data is a direct consequence of the severe experimental challenges inherent to hydroacoustic measurements at high Reynolds numbers ($Re_L > 10^7$) \citep{Joseph2017Pressure,Ciappi2005characteristics}: (i) Background noise contamination: Achieving operationally relevant Reynolds numbers requires high freestream velocities, which increases facility acoustic noise generated by propulsion fans and flow-wall interactions. This background noise, propagating primarily as low-frequency plane waves, can spectrally mask the hydrodynamic pressure signals of interest, resulting in a critically low signal-to-noise ratio that may mask large-scale turbulent motions. (ii) Sensor spatial resolution limits: As the Reynolds number increases, the viscous length scale ($\nu/u_\tau$) decreases drastically, shrinking the physical size of the high-frequency turbulent eddies. When these small-scale structures become smaller than the sensing area of conventional microphones, the sensor integrates pressure over its face rather than at a point. This spatial averaging effect causes a severe artificial attenuation of the high-frequency spectral tail, preventing the accurate resolution of the fine-scale turbulence. (iii) Complex flow interference by appendages: Unlike canonical flat plates, the fully appended geometry introduces strong coherent structures, most notably the unsteady horseshoe vortex systems at appendage-hull junctions. These energetic vortices induce massive, localized pressure fluctuations that interact non-linearly with the developing hull boundary layer. Distinguishing these appendage-induced disturbances from the underlying boundary layer turbulence requires precise sensor placement and interpretation, further complicating the measurement campaign. Consequently, establishing a high-quality database that overcomes these challenges, through advanced signal processing and sensor techniques, to achieve high Reynolds numbers, low background noise, and high spatial resolution, remains a critical objective.}

The presence of appendages, notably the sail (fairwater) and stern control surfaces, serves to intensify the aforementioned non-equilibrium effects, thereby imposing a substantial hydrodynamic and acoustic penalty, as outlined in  \citet{Jiang2025PoFSUBOFF}. These appendages act as active generators of coherent vortical structures \citep{zhang2020review}. Moreover, the adverse pressure gradient induced at the appendage-hull leading edge forces the incoming boundary layer to roll up, forming a highly unsteady horseshoe vortex (HSV) system. This energetic vortex system wraps around the appendage and convects downstream, effectively creating strong pressure fluctuations that impinge directly on the stern and the propeller inflow. Recent investigations explicitly link this non-uniform wake field to severe propeller vibration and radiated noise issues \citep{liu2023control, Wu2023VCBCFD}. Consequently, the suppression of these junction flows must be considered not only from the perspective of hydrodynamic efficiency, but also from that of acoustic requirements. Traditional strategies have relied on local geometric optimization, such as optimizing the sail-hull junction with fillets \citep{toxopeus2014improvement, rahmani2023experimental, wang2021numerical} or refining the sail trailing edge profile \citep{ma2025performance}. However, these methods often yield limited success in breaking the highly coherent HSV cores. To overcome this limitation, this study adopts the vortex control baffle, a concept originally pioneered by Liu \textit{et al.} \citep{Liu2010Numerical, Liu2014Method} to physically break the junction vortex core. While initial applications focused on improving wake uniformity and propulsion efficiency \citep{Liu2010Numerical}, subsequent numerical investigations by \citet{Wu2023VCBCFD} and \citet{liu2023control} have optimized the baffle's geometry and considered its potential for vibration reduction. Building upon these foundations, yet distinct from purely numerical or wake-focused studies, we present the first experimental study of the VCB's efficacy in suppressing wall-pressure fluctuations. By repurposing this mechanism to disrupt the HSV coherence at its source, this approach offers a promising, as yet unproven, strategy for the reduction of global noise.

Beyond the straight-ahead condition typically addressed previously, the operational conditions of a submarine involve complex maneuvering states, which fundamentally reshape the flow topology. As demonstrated in the important works of  \citet{ashok2015asymmetries, ashok2015structure}, maneuvers such as yaw and pitch transform the axisymmetric wake into a complex system dominated by asymmetric streamwise vortices. During  maneuvers, the submarine model effectively acts as a low-aspect-ratio lifting body, inducing strong crossflow pressure gradients that drive three-dimensional separation and reattachment. However, a critical disparity exists in the literature: while the wake properties \citep{ashok2013turbulent} and integrated hydrodynamic loads \citep{leong2016evaluation} have been characterized, the response of the unsteady wall-pressure field remains unexplored. This omission is significant because the pressure response is expected to be highly non-linear and spatially dependent. Preliminary investigations suggest a bifurcation in flow regimes, where leeward separation bubbles may shield the wall from high-frequency turbulence, whereas windward vortex impingement massively amplifies broadband noise. Understanding these spatially evolving, non-monotonic mechanisms is crucial for defining the full maneuvering acoustic properties of submarines, yet systematic experimental data at high Reynolds numbers remain absent.

This paper addresses these critical gaps through a systematic wind tunnel investigation of the fully appended SUBOFF model at high Reynolds numbers ($Re_L$ up to $1.4 \times 10^7$). To ensure the fidelity of the wall-pressure measurements in this challenging environment, the study employs high-frequency pinhole microphones coupled with comprehensive signal processing techniques to resolve the true hydrodynamic pressure spectrum free from facility acoustic contamination. The specific objectives of this work are threefold: (i) To establish a benchmark database of wall-pressure fluctuations for the fully appended hull, rigorously quantifying the spectral amplification induced by the sail and stern fins relative to the bare hull baseline. (ii) To experimentally study the vortex control baffle as a novel noise suppression device. This study provides the first experimental assessment of its ability to physically disrupt the coherent sail-junction horseshoe vortex at its source, thereby suppressing the downstream wake impingement and global pressure fluctuations. (iii) To investigate the complex spatio-temporal evolution of the pressure field under static maneuvering conditions (yaw and pitch). This includes investigating the non-linear physics of crossflow separation, vortex-wall interactions, and the bifurcation of flow regimes on leeward and windward sides. By integrating high-Reynolds-number experimentation with the flow control and maneuvering analysis, this study provides new physical insights into the flow noise on complex underwater vehicles.

The remainder of this paper is organized as follows. \S \ref{Experimental_Methodology} details the experimental methodology, describing the high-speed wind tunnel facility, the appended SUBOFF model geometry, and the sensor instrumentation employed for both mean and fluctuating pressure measurements. This \S also outlines the data acquisition and multi-stage signal processing strategy used to correct for pinhole resonance and background noise. \S \ref{Results_and_Discussion} presents the comprehensive analysis of the experimental data. This begins with a validation of the mean flow and a characterization of the baseline wall-pressure spectra under straight-ahead conditions in \S \ref{flow_overview}, examining the influence of Reynolds number and spectral properties. Subsequently, \S \ref{Effects_of_appendages} quantifies the severe amplification of wall-pressure fluctuations caused by appendages, identifying the dominant noise sources. \S \ref{Effects_of_vortex_control_baffle} evaluates the effectiveness of the vortex control baffle, demonstrating its mechanism in suppressing large-scale coherent structures. The study then investigates the complex non-monotonic evolution of the pressure field under maneuvering conditions (yaw and pitch) in \S \ref{Effects_of_Maneuvering_Conditions}. Finally, \S \ref{Conclusions} summarizes the key findings of this work and their implications for hydroacoustic design, along with suggestions for future research.

\section{Experimental Methodology}\label{Experimental_Methodology}

\subsection{Wind Tunnel Facility and Test Model}\label{Wind_Tunnel_SUBOFF_Model}

The experiments were performed in the high-speed test section ($3 \text{ m} \times 2.5 \text{ m}$ in cross section) of the Wind Tunnel in the State Key Laboratory of Ocean Engineering, Shanghai Jiao Tong University. Although the full-scale vehicle operates in water, wind tunnel testing was selected to achieve the requirement of high Reynolds numbers ($O(10^7)$) and to allow the use of high-frequency precision microphones without the complexities of waterproofing. This approach relies on Reynolds number similarity, a methodology validated in our prior work \citep{Jiang2025HullExperiment} and widely accepted in the literature \citep{Goody1999Experimental, Meyers2015Wallpressure, Joseph2020JFMPlate, Gibeau2021JFMPlate}.

\begin{figure}
	\centering
	\begin{minipage}[c]{1\linewidth}
    	\centering
		\begin{overpic}[width=12cm]{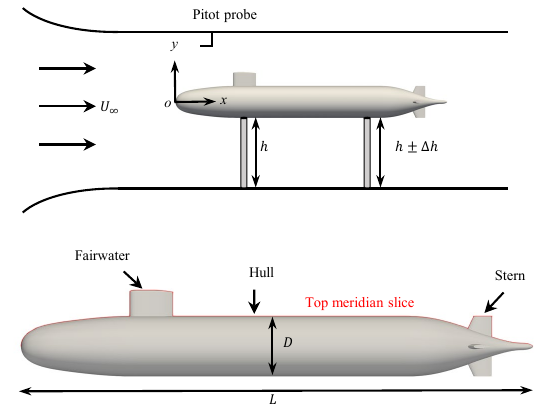}
			\put(-5,75){\color{black}{(a)}}
            \put(-5,25){\color{black}{(b)}}
		\end{overpic}
	\end{minipage}
    \caption{Schematic of the appended SUBOFF model experimental configuration: (a) Overall installation of the SUBOFF model in the high-speed test section of the Shanghai Jiao Tong University wind tunnel; (b) Detailed view of the geometrical information of the test model.}
    \label{fig:windtunnel_arrangement}
\end{figure}

\begin{figure}
	\centering
	\begin{minipage}[c]{1\linewidth}
    	\centering
		\begin{overpic}[width=13.5cm]{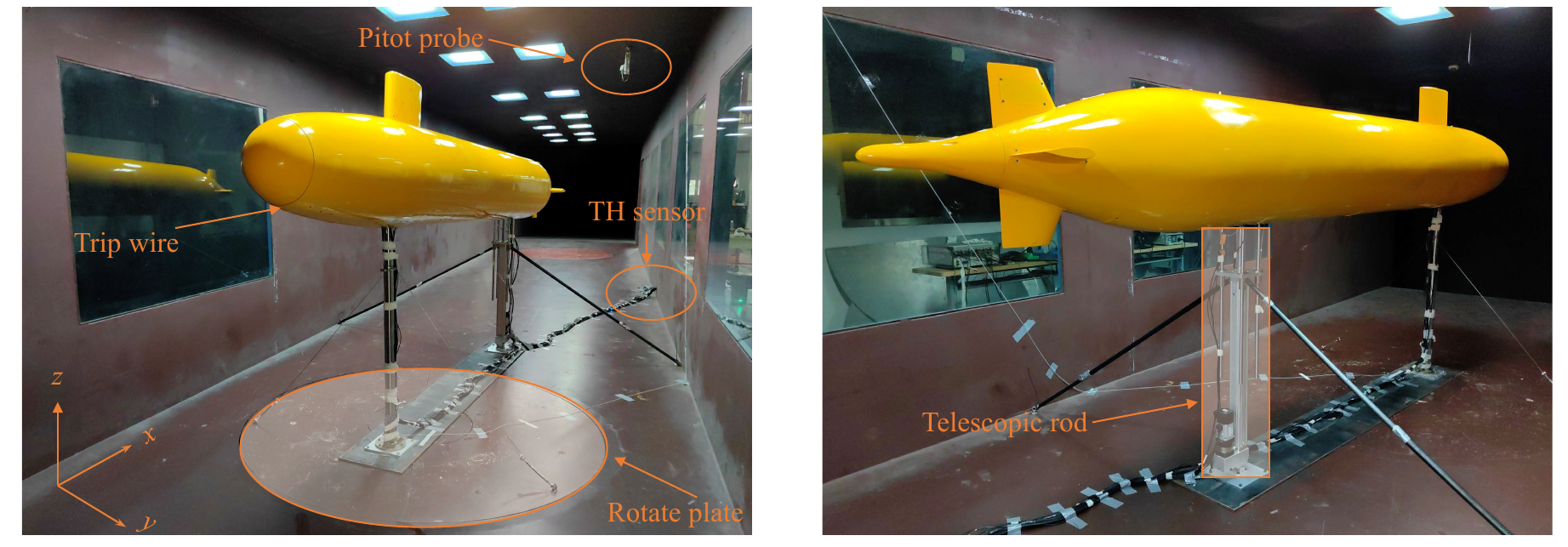}
			\put(-2,33){\color{black}{(a)}}
            \put(49,33){\color{black}{(b)}}
		\end{overpic}
	\end{minipage}
    \caption{Photographs illustrating the SUBOFF model with appendages installed in the wind tunnel test section: (a) View from upstream (looking downstream) showing the model's overall placement and the flow direction; (b) View from downstream (looking upstream) detailing the stern fins.}
    \label{fig:Exp_photo}
\end{figure}

This study utilizes the 1:24 scale DARPA SUBOFF model \citep{groves1989geometric}, a canonical benchmark for submarine hydrodynamics. The model geometric parameters are summarized in Table~\ref{tab:suboff_parameters}. The model features an overall length of $L=4.356$\,m and a maximum diameter of $D=0.508$\,m, resulting in a blockage ratio of 2.70\% in the present wind tunnel, which minimizes wall interference effects. To trigger a fully developed turbulent boundary layer, a 0.50\,mm diameter trip wire was azimuthally mounted at $0.75D$ downstream of the nose, consistent with established wake studies \citep{Jimenez2010JFMwakeflow}. \textcolor{black}{In addition, to ensure the boundary layers over the appendages were fully developed, a trip wires were also applied to the sail and stern appendages at 5\% of their chord lengths from the leading edge, following previous experimental and numerical recommendations \citep{Liu1998SUBOFFexp, MorseMahesh2023Trippingeffects}.} The model was supported by two streamlined struts positioned downstream of the measurement region to prevent support interference on the developing boundary layer (see Figure~\ref{fig:Exp_photo}). A primary objective of this study is to assess the impact of appendages and flow control devices on wall-pressure fluctuations relative to the bare-hull baseline \citep{Jiang2025HullExperiment}. To achieve this, comparative measurements were performed across three distinct model configurations: the baseline bare hull, the fully appended SUBOFF, and the appended SUBOFF equipped with vortex control baffles, as detailed in Table~\ref{tab:model_configurations}. The complete experimental arrangement is depicted in Figures~\ref{fig:windtunnel_arrangement} and \ref{fig:Exp_photo}, with sensor coordinates detailed in Table~\ref{tab:sensor_positions}.

\begin{table}
\centering
\def~{\hphantom{0}}
\begin{tabular}{lcc}
\multicolumn{2}{c}{Generic submarine type} & The appended SUBOFF model\\ 
Description& Symbol& Magnitude\\ 
Length overall& $L$& 4.356 $\mathrm{m}$\\
Maximum hull diameter& $D$& 0.508 $\mathrm{m}$\\
Volume of displacement& $\Delta$& 0.718 $\mathrm{m^3}$\\
Wetted surface area& $S$& 6.338 $\mathrm{m^2}$\\ 
\end{tabular}
\caption{Main geometrical parameters of the appended SUBOFF model.}
\label{tab:suboff_parameters}
\end{table}

\begin{table}
\centering
\def~{\hphantom{0}}
\begin{tabular}{l p{10cm}}
Configuration & Description \\
(i) Bare Hull & The baseline axisymmetric body of revolution without any appendages. Detailed geometric parameters are provided in \citet{Jiang2025HullExperiment}. \\
\addlinespace
(ii) Appended SUBOFF & The fully appended model featuring the sail (fairwater) positioned on the top meridian and four identical stern appendages arranged in a cruciform configuration (see Figure~\ref{fig:Exp_photo}). \\
\addlinespace
(iii) Appended + VCB & The appended SUBOFF model equipped with vortex control baffles at the sail-hull and fin-hull junctions (see Figure~\ref{fig:Schematic_of_VCB}). The VCB is designed to disrupt the coherent horseshoe vortex structure (detailed geometry in \S \ref{Effects_of_vortex_control_baffle}). \\
\end{tabular}
\caption{Summary of the experimental model configurations.}
\label{tab:model_configurations}
\end{table}

\subsection{Instrumentation and Data Acquisition}\label{Instrumentation_Data_Acquisition}
\begin{figure} 
	\centering
	\begin{minipage}[c]{1\linewidth}
    	\centering
		\begin{overpic}[width=13cm]{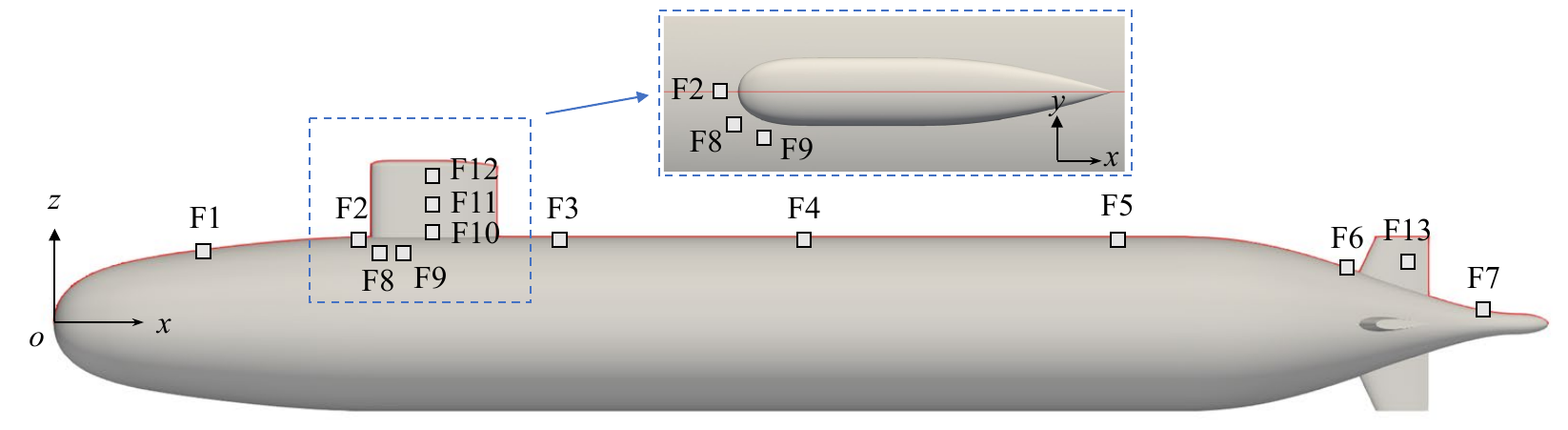}
		\end{overpic}
	\end{minipage}
    \caption{Schematic of the arrangement of measurement points along the test model top meridian line. Square symbols denote wall-pressure fluctuation sensors.}
    \label{fig:SensorArrangement}
\end{figure}

\begin{table}
\centering
\begin{tabular}{l c c c l}
Sensor Type & Sensor ID & Axial Position \( x \) (m) & Normalized \( x/L \) & Location / Description \\ 
\multirow{8}{*}{\centering\parbox{2cm}{\centering Static \\ pressure}} 
  & S1 & 0.788 & 0.181 & Nose \\
  & S2 & 1.751 & 0.402 & Middle section \\
  & S3 & 2.618 & 0.601 & Middle section \\
  & S4 & 3.228 & 0.741 & Middle section \\
  & S5 & 3.441 & 0.790 & Aft section \\
  & S6 & 3.659 & 0.840 & Aft section \\
  & S7 & 4.038 & 0.927 & Tail \\
  & S8 & 4.260 & 0.978 & Tail end \\ 
  \hline
\multirow{13}{*}{\centering\parbox{2cm}{\centering Fluctuation \\ pressure}} 
 & F1 & 0.457 & 0.105 & Hull: Behind trip wire \\
 & F2 & 0.904 & 0.207 & Hull: Sail leading edge \\
 & F3 & 1.481 & 0.340 & Hull: Middle section \\
 & F4 & 2.182 & 0.501 & Hull: Middle section (ZPG ref.) \\
 & F5 & 3.092 & 0.710 & Hull: Aft section \\
 & F6 & 3.795 & 0.871 & Hull: Fin leading edge \\
 & F7 & 4.164 & 0.956 & Hull: Tail end \\ \cmidrule{2-5}
 & F8 & 0.936 & 0.215 & Sail Junction Side ($z=44$\,mm) \\
 & F9 & 1.002 & 0.229 & Sail Junction Side ($z=53$\,mm) \\ \cmidrule{2-5}
 & F10 & 1.109 & 0.255 & Sail Surface: $0.5c$, $0.1h$ (Root) \\
 & F11 & 1.109 & 0.255 & Sail Surface: $0.5c$, $0.5h$ (Mid-span) \\
 & F12 & 1.109 & 0.255 & Sail Surface: $0.5c$, $0.9h$ (Tip) \\
 & F13 & 3.931 & 0.902 & Stern Fin Surface Center \\ 
\end{tabular}
\caption{Positions of static and surface pressure sensors along the SUBOFF model. For appendage sensors, specific local locations are noted in the description, where $c = 368.29$\,mm and $h = 205.57$\,mm denote the chord length and span height of the sail, respectively.}
\label{tab:sensor_positions}
\end{table}

A comprehensive instrumentation setup was employed to characterize both the mean flow topology and the unsteady wall-pressure field. To capture the complex flow physics induced by the appendages, the array of wall-pressure fluctuation sensors was expanded beyond the baseline hull configuration (F1--F7). Specifically, two sensors (F8, F9) were installed in the immediate vicinity of the sail leading-edge junction to resolve the unsteady dynamics of the horseshoe vortex system. To characterize the boundary layer development and potential flow separation on the appendages, three sensors (F10--F12) were mounted on the lateral surface of the sail at 50\% chord length ($0.5c$), distributed spanwise at 10\%, 50\%, and 90\% of the sail height, respectively. Additionally, a sensor (F13) was flush-mounted on the lateral surface of the top stern fin to monitor the wake flow characteristics in the aft region. The positions of all fluctuation sensors are summarized in Table~\ref{tab:sensor_positions}, with the sensor arrangement along the top meridian line depicted in Figure~\ref{fig:SensorArrangement}. The static pressure distribution was acquired using eight pressure taps (S1--S8) connected to high-precision differential pressure transducers via 2\,mm inner-diameter flexible tubing. The transducers possess an accuracy of $\pm 0.1\%$ full scale. Data were sampled at 1 kHz for sufficient duration to ensure statistical convergence of the mean flow properties. Fluctuating wall pressures were resolved using 13 flush-mounted microphones. These sensors feature a measurement range of $\pm 2$\,kPa and a sensitivity of 2.5\,mV/Pa. Crucially, to reduce spatial averaging effects which attenuate high-frequency energy in TBL measurements \citep{Gravante1998dPlus}, each microphone was fitted with a pinhole cap (diameter $d = 1.2$ mm).
Data acquisition was performed using a National Instruments PXIe system (NI-395) with a 24-bit resolution. Signals were sampled simultaneously at a frequency of $f_s = 15$\,kHz for a duration of $T = 90$\,s per test case, ensuring sufficient resolution of the spectral content and convergence of lower-order statistics.

\begin{table}
	\centering
    \def~{\hphantom{0}}
	\begin{tabular}{ccc}
		$U_{\infty}$ (m/s) & $d^+$ Range & Estimated $f_{\text{max}}$ Range (kHz) \\ 
		20                         & $20 \sim 60$         & $11.9 \sim 35.6$                                \\
		30                         & $30 \sim 90$         & $17.8 \sim 53.5$                                \\
		43                         & $40 \sim 120$        & $23.8 \sim 71.3$                                \\
		50                         & $50 \sim 140$        & $29.7 \sim 83.2$                                \\ 
	\end{tabular}
	\caption{Summary of non-dimensional sensor diameters ($d^+$) and estimated maximum resolvable frequencies ($f_{\text{max}}$) for different flow speeds. The $f_{\text{max}}$ values correspond to a 2 dB attenuation limit, estimated using the relation $f_{\text{max}} \approx 57.2 u_{\tau} / d$ based on approximations from \citet{Gravante1998dPlus} and \citet{Meyers2015Wallpressure}.}
	\label{tab:sensor_resolution}
\end{table}

\textcolor{black}{Another significant consideration is that the finite spatial extent of the wall-pressure sensors can lead to the attenuation of high-frequency spectral components due to spatial averaging. To assess this effect, the non-dimensional sensing diameter, $d^+ = d u_{\tau} / \nu$, was calculated along the hull surface, where $d = 1.2$ mm is the sensor diameter. The local friction velocity, $u_{\tau}$, was obtained from RANS simulations that were configured to strictly replicate the flow conditions and geometry of the wind tunnel experiments. The calculated $d^+$ values for all investigated flow speeds are summarized in Table \ref{tab:sensor_resolution}. It is noted that these values generally exceed the conservative threshold of $d^+ \le 18$ suggested by \citet{Gravante1998dPlus} to completely avoid spectral attenuation. However, the critical parameter for the present study is the physical frequency limit imposed by the sensor size. Following the methodology described by \citet{Meyers2015Wallpressure} and \citet{Balantrapu2023APG}, the maximum resolvable frequency (corresponding to a 2 dB attenuation) can be estimated. \citet{Gravante1998dPlus} observed a 2 dB attenuation at a non-dimensional frequency of $f^+_{2\text{dB}} \approx 2.2$ for a sensor size of $d^+ \approx 26$. Assuming that the cutoff frequency scales inversely with the sensing diameter ($f^+ d^+ \approx \text{const}$), the maximum useful physical frequency can be estimated as $f_{\text{max}} \approx 57.2 u_{\tau} / d$ \citep{Meyers2015Wallpressure,Joseph2020JFMPlate,Balantrapu2023APG}. As shown in Table \ref{tab:sensor_resolution}, the lowest predicted cutoff frequency occurs at the lowest flow speed ($U_{\infty} = 20$ m/s) in regions of minimal wall shear stress, yielding a conservative limit of approximately 11.9 kHz. At higher flow speeds, although $d^+$ values increase significantly (up to 140 at 50 m/s), the physical frequency limits also increase (up to 83.2 kHz) due to the higher friction velocities. Since the current analysis is strictly focused on a frequency range up to 7.5 kHz, the estimated sensor resolution limits for all cases are well above the frequency of interest. Therefore, the measured spectra are considered to be free from significant spatial averaging effects within the analyzed bandwidth, and no spectral corrections were applied.}

\subsection{Signal Processing Overview}\label{Signal_Processing_main}
To ensure the high-fidelity resolution of hydrodynamic wall-pressure fluctuations and to minimize facility-related contamination, a rigorous multi-stage signal processing pipeline was implemented. This procedure primarily addresses three experimental challenges: (i) the Helmholtz resonance induced by the microphone pinhole geometry, (ii) low-frequency acoustic background noise from the wind tunnel propulsion system, and (iii) the statistical convergence of the power spectral density estimates. 

The correction methodology is based on established best practices for turbulent boundary layer measurements \citep{Joseph2017Pressure, Baars2024JFMPlate} and follows the validated framework from our previous studies \citep{Jiang2025HullExperiment}. In brief, dynamic transfer functions were applied to deconvolve the sensor resonance, while a Wiener filter was employed to isolate the hydrodynamic signals from plane-wave acoustic noise. For a comprehensive description of the mathematical models, calibration procedures, and the rigorous validation of this methodology against canonical flat-plate benchmarks, the reader is referred to Appendix~\ref{Signal_Processing}.

\subsection{Experimental Conditions}\label{Summary_data_set}

\begin{table}
\centering
\def~{\hphantom{0}}
\begin{tabular}{c c c c c c }
\multicolumn{2}{c}{Parameter} & Case 1 & Case 2 & Case 3 & Case 4  \\
Flow Velocity & $U_{\text{air}}$ ($\mathrm{m/s}$)  & 20 & 30 & 43 & 50 \\
Reynolds Number & $Re_L$ ($\times 10^6$) & 5.6 & 8.4 & 12.0 & 14.0 \\
\end{tabular}
\caption{Summary of air velocity ($U_{\text{air}}$) and corresponding Reynolds number ($Re_L$) for the experimental test cases. (Calculated based on the model length $L = 4.356\,\text{m}$ and kinematic viscosity $\nu = 1.56\times10^{-5}\,\text{m}^2/\text{s}$).}
\label{tab:reynolds_relation}
\end{table}

The experimental conditions covered a range of Reynolds numbers, maneuvering attitudes, and control devices to investigate wall pressure fluctuations systematically. The test matrix is defined by:
\begin{enumerate}[label=(\roman*)]
    \item Reynolds Number ($Re_L$): Tests were conducted at freestream velocities of $U_{\infty} = 20, 30, 43, \text{ and } 50$\,m/s, corresponding to length-based Reynolds numbers ranging from $5.6 \times 10^6$ to $1.4 \times 10^7$ (see Table~\ref{tab:reynolds_relation}).
    \item Maneuvering Conditions: To simulate operational maneuvers, the model was tested at static Yaw angles ($ 0^\circ, 3^\circ, 6^\circ$) and Pitch angles ($0^\circ, \pm 3^\circ, \pm 6^\circ$).
\end{enumerate}

\section{Results and Discussion}\label{Results_and_Discussion}
\textcolor{black}{This section presents a systematic analysis of the experimental findings, structured to elucidate the physics of wall-pressure fluctuations on the fully appended submarine model. 
First, the fundamental flow topology is characterized to validate the experimental setup and establish a hydrodynamic baseline. 
Subsequently, the study delves into the spectral characteristics of the wall-pressure fluctuations, with a particular focus on the noise amplification mechanisms induced by the sail and stern appendages. Finally, the effects of complex maneuvering attitudes are examined, followed by a quantitative evaluation of the noise suppression efficacy of the vortex control baffle.}

\subsection{Flow Characteristics of the Test Model}\label{flow_overview}

\textcolor{black}{Before quantifying the unsteady wall-pressure fluctuations (the primary noise source), it is essential to establish the global characteristics of the flow field. 
Firstly, this subsection evaluates the baseline flow field through mean pressure distributions and boundary layer parameters. 
Understanding these mean flow properties, particularly the locations of adverse pressure gradients and flow separation, which provides the necessary physical context for interpreting the spectral behaviors and scaling laws discussed in subsequent sections.}

\subsubsection{Mean Flow Characteristics}\label{Mean_Flow}
\begin{figure}
	\centering
	\begin{overpic}[width=10cm]{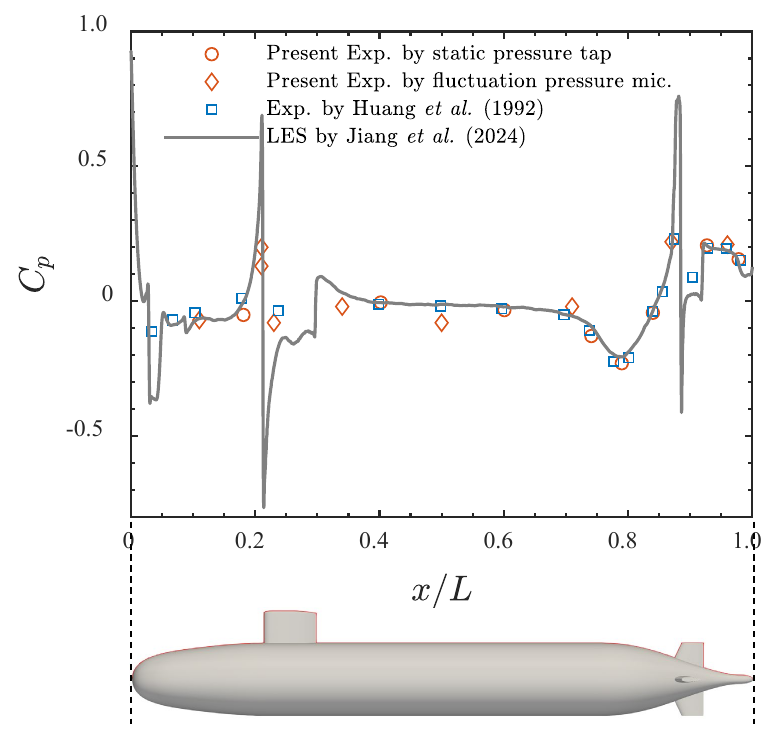}
	\end{overpic}
	\caption{Mean pressure coefficient ($C_p$) distribution along the top meridian of the appended SUBOFF model at $Re = 1.2 \times 10^7$ (straight-ahead condition) compared with experimental data from \citet{Liu1998SUBOFFexp} and WMLES from \citet{Jiang2025PoFSUBOFF}.}
	\label{fig:AFF8_0deg_43_mean_pressure}
\end{figure}
To characterize the baseline flow field and validate the fidelity of the experimental setup, the mean pressure coefficient ($C_p$) distribution along the top meridian line of the SUBOFF hull was measured. The pressure coefficient is defined as $C_p = (\bar{p} - p_{\infty})/\frac{1}{2} \rho U_{\infty}^2$, where $\bar{p}$ is the local time-averaged static pressure, $p_{\infty}$ is the freestream static pressure, $\rho$ is the fluid density, and $U_{\infty}$ is the freestream velocity. Figure~\ref{fig:AFF8_0deg_43_mean_pressure} presents the $C_p$ distribution at a Reynolds number of $Re = 1.2 \times 10^7$ (corresponding to $U_{\infty} = 43$ m/s) under the straight-ahead condition. The present experimental data, acquired from both the static pressure taps and the mean component of the fluctuation microphones, are compared against the benchmark experimental data of  \citet{Liu1998SUBOFFexp} and the high-fidelity Wall-Modeled Large Eddy Simulation (WMLES) results of \citet{Jiang2025PoFSUBOFF}. The results demonstrate an excellent agreement between the present measurements and the reference datasets across the entire model length.

The flow exhibits three distinct regimes characterized by the model geometry:
(i) Forebody ($x/L < 0.2$): The flow accelerates rapidly over the nose, creating a suction peak, followed by a sharp pressure variation near $x/L \approx 0.2$. This fluctuation corresponds to the stagnation and subsequent acceleration induced by the leading edge of the sail (fairwater), confirming that the sensors accurately capture the appendage-induced pressure gradients.
(ii) Parallel mid-body ($0.3 < x/L < 0.7$): In this region, the $C_p$ curve remains relatively flat and close to zero. This zero-pressure-gradient condition is critical for establishing an equilibrium turbulent boundary layer, which serves as the baseline for the subsequent fluctuation analysis.
(iii) Stern ($x/L > 0.7$): The flow undergoes pressure recovery characterized by strong adverse pressure gradients followed by a favorable gradient near the tail tip. The rapid pressure variations near the stern ($x/L \approx 0.9$) are attributed to the interference effects of the fins. The capability of the measurement system to accurately resolve these steep pressure gradients generated by the appendages confirms the fidelity of the current experimental methodology. Notably, the consistency between the static pressure taps and the mean values extracted from the fluctuation microphones provides a robust validation of the microphone calibration. The agreement with established benchmarks confirms that the experimental facility and model installation replicate the canonical flow physics of the SUBOFF hull, providing a reliable foundation for the analysis of wall-pressure fluctuations.

\subsubsection{Wall-Pressure Spectra Analysis}\label{Wall-pressure_spectra}
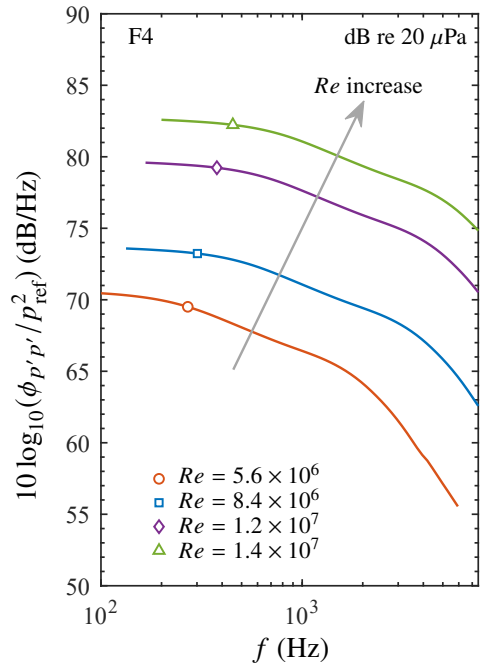
\begin{figure}
\centering
	\begin{minipage}[c]{1\linewidth}
     \centering
	\begin{overpic}[width=5.5cm]{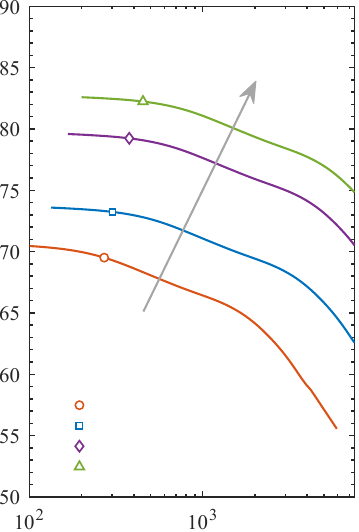}
        \put(-9,20){{\color{black}{\rotatebox{90}{$10 \log_{10}(\phi_{p^\prime p^\prime}/p^2_\text{ref})\, \text{(dB/Hz)}$}}}}
        \put(30,-5){{\color{black}{$f$ (Hz)}}}
        \put(10,94){\color{black}\small{F4}}
        \put(45,94){\color{black}\footnotesize{dB re 20 $\mu$Pa}}
        \put(40,86){\color{black}\footnotesize{$Re$ increase}}
        \put(18,23){\color{black}\footnotesize{$Re = 5.6 \times 10^6$}}
        \put(18,19){\color{black}\footnotesize{$Re  = 8.4 \times 10^6$}}
        \put(18,15){\color{black}\footnotesize{$Re = 1.2 \times 10^7$}}
        \put(18,11){\color{black}\footnotesize{$Re = 1.4 \times 10^7$}}
	\end{overpic}
	\end{minipage}
    \vspace{0.5cm}
    \caption{Power spectral density of wall-pressure fluctuations measured at sensor F4 (parallel mid-body, $x/L = 0.5$) on the fully-appended hull, shown for a range of Reynolds numbers ($Re$) under straight-ahead conditions.}
    \label{fig:F4_Impact_of_Re}
\end{figure}

\paragraph{\textcolor{black}{Reynolds number scaling and baseline selection:}} To ensure the generality of the results, the sensitivity of wall-pressure spectra to Reynolds number variations is first evaluated. Figure~\ref{fig:F4_Impact_of_Re} illustrates the power spectral density measured at the reference sensor F4 ($x/L = 0.5$) across Reynolds numbers from $Re = 5.6 \times 10^6$ ($U_{\infty}=20$ m/s) to $1.4 \times 10^7$ ($U_{\infty}=50$ m/s). The spectra exhibit remarkable self-similarity. With increasing $Re$, the PSD curves shift upwards, strictly following the dynamic pressure scaling ($p'_{rms} \propto q_\infty \propto U_\infty^2$). This confirms that the measurements are capturing hydrodynamic pressure fluctuations rather than acoustic noise.

Based on these observations, the condition of $U_{\infty} = 20$ m/s ($Re = 5.6 \times 10^6$) is selected as the baseline for the subsequent detailed spatial analysis. This selection is justified by three key factors: (i) Bandwidth and sampling constraints: Due to the present number of simultaneous measurement channels, the data acquisition system imposed a maximum sampling rate of $f_s = 15$ kHz per channel, corresponding to a Nyquist frequency of $f_{Nyq} = 7.5$ kHz. Since turbulent characteristic frequencies scale with velocity ($f \propto U_{\infty}$), higher flow speeds would shift significant spectral energy towards higher frequencies, potentially exceeding the available bandwidth, as evident in Figure~\ref{fig:F4_Impact_of_Re}. The 20 m/s condition ensures that the energy-containing eddies and the inertial subrange are fully resolved within the 7.5 kHz analysis bandwidth, avoiding high-frequency cutoff. (ii) Signal-to-noise ratio (SNR): The $20$ m/s condition offers an optimal balance where the hydrodynamic pressure signal is sufficiently strong, while the background facility noise (which scales non-linearly with speed) remains low, resulting in a superior SNR, particularly in the low-frequency range. (iii) Representativeness: Given the demonstrated self-similarity of the spectra in Figure~\ref{fig:F4_Impact_of_Re}, the flow physics captured at 20 m/s are representative of the higher Reynolds number cases, allowing for generalized conclusions regarding spatial evolution and appendage effects.

\begin{figure} 
    \centering
    \vspace{0.1cm}
	\begin{minipage}[c]{1.0\linewidth}
    	\centering
    \begin{overpic}[width=4.0cm]{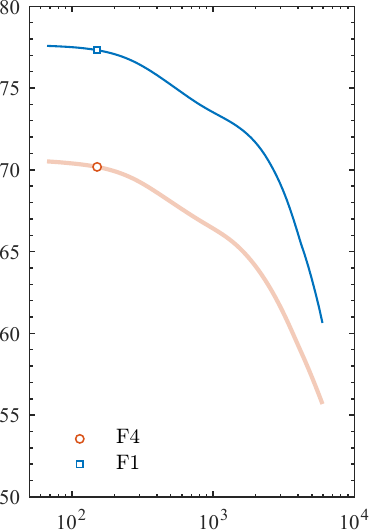}
        \put(-9,98){\color{black}{(a)}}
        \put(-9,20){\small{\color{black}{\rotatebox{90}{$10 \log_{10}(\phi_{p^\prime p^\prime}/p^2_\text{ref})\, \text{(dB/Hz)}$}}}}
        \put(29,-5){\small{\color{black}{$f$ (Hz)}}}
    \end{overpic}
    \hspace{0.4cm}
    \begin{overpic}[width=4.0cm]{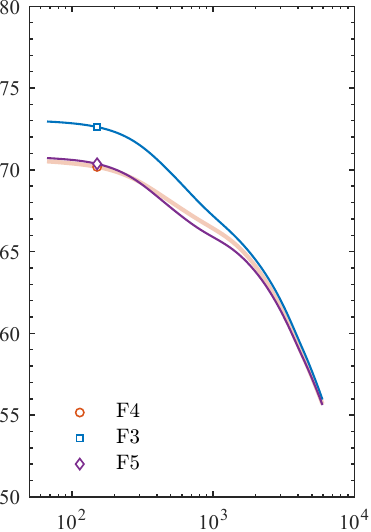}
        \put(-9,98){\color{black}{(b)}}
        \put(29,-5){\small{\color{black}{$f$ (Hz)}}}
    \end{overpic}
    \hspace{0.4cm}
    \begin{overpic}[width=4.0cm]{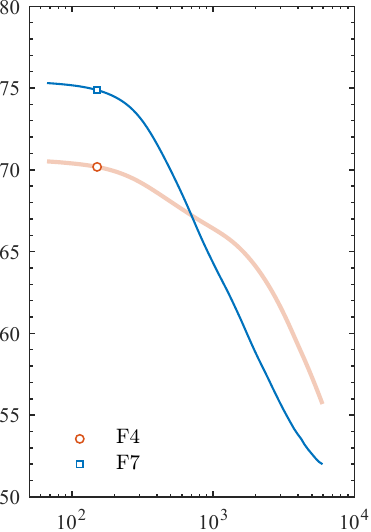}
        \put(-9,98){\color{black}{(c)}}
        \put(29,-5){\small{\color{black}{$f$ (Hz)}}}
    \end{overpic}
    \end{minipage}
    \vspace{0.5cm}
    \caption{Wall-pressure fluctuation PSD at $Re = 5.6 \times 10^6$  for the straight-ahead condition, showing the effect of axial position. The spectrum at F4 ($x/L=0.5$) is used as the baseline for comparison with: (a) the developing flow region (F1); (b) other parallel mid-body locations (F3, F5); (c) the aft-body stern region (F7).}
    \label{fig:AFF8_hull_PSD}
\end{figure}

\begin{figure}
\centering
	\begin{minipage}[c]{1\linewidth}
     \centering
	\begin{overpic}[width=4.5cm]{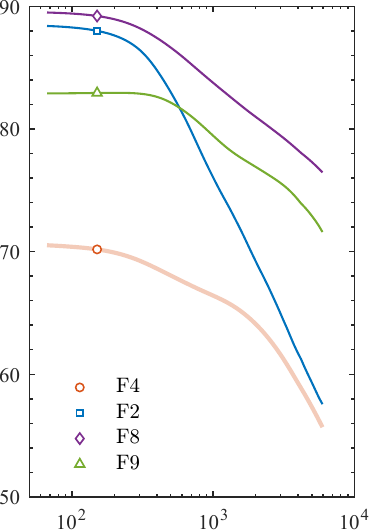}
        \put(-9,98){\color{black}{(a)}}
        \put(-9,20){{\color{black}{\rotatebox{90}{$10 \log_{10}(\phi_{p^\prime p^\prime}/p^2_\text{ref})\, \text{(dB/Hz)}$}}}}
        \put(30,-5){{\color{black}{$f$ (Hz)}}}
	\end{overpic}
    \hspace{0.5cm}
    \begin{overpic}[width=4.5cm]{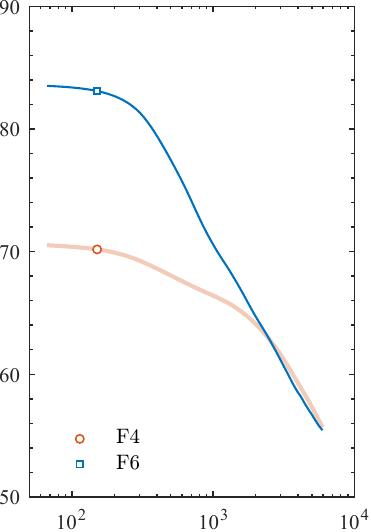}
        \put(-9,98){\color{black}{(b)}}
        \put(30,-5){{\color{black}{$f$ (Hz)}}}
    \end{overpic}
	\end{minipage}
    
    \vspace{0.5cm} 
    
	\begin{minipage}[c]{1\linewidth}
     \centering

	\begin{overpic}[width=4.5cm]{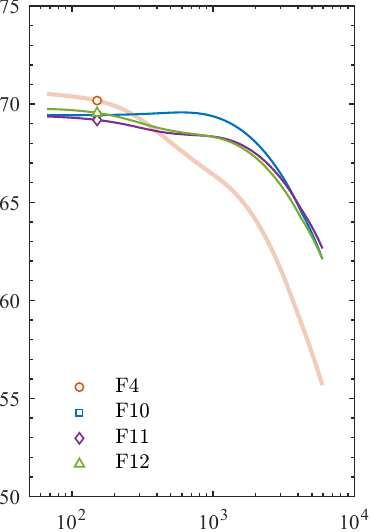}
        \put(-9,98){\color{black}{(c)}} 
        \put(-9,20){{\color{black}{\rotatebox{90}{$10 \log_{10}(\phi_{p^\prime p^\prime}/p^2_\text{ref})\, \text{(dB/Hz)}$}}}}
        \put(30,-5){{\color{black}{$f$ (Hz)}}}
	\end{overpic}
    \hspace{0.5cm}
    \begin{overpic}[width=4.5cm]{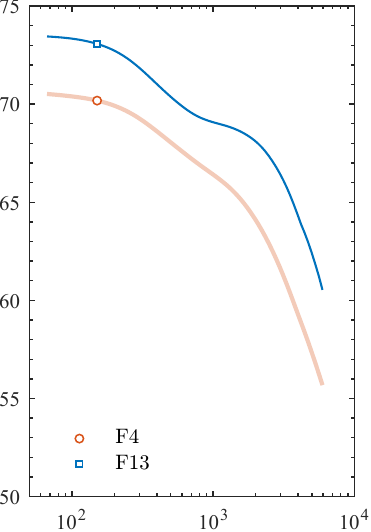}
        \put(-9,98){\color{black}{(d)}} 
        \put(30,-5){{\color{black}{$f$ (Hz)}}}
    \end{overpic}
	\end{minipage}
    \vspace{0.5cm}
    \caption{Comparison of wall-pressure PSD around appendages against the mid-body baseline (F4) at $Re = 5.6 \times 10^6$. (a) Sail-hull junction (F2, F8, F9); (b) Stern-fin junction (F6); (c) Sail surface (F10--F12); (d) Stern fin surface (F13).}
    \label{fig:AFF8_Combined_Appendage_PSD}
\end{figure}
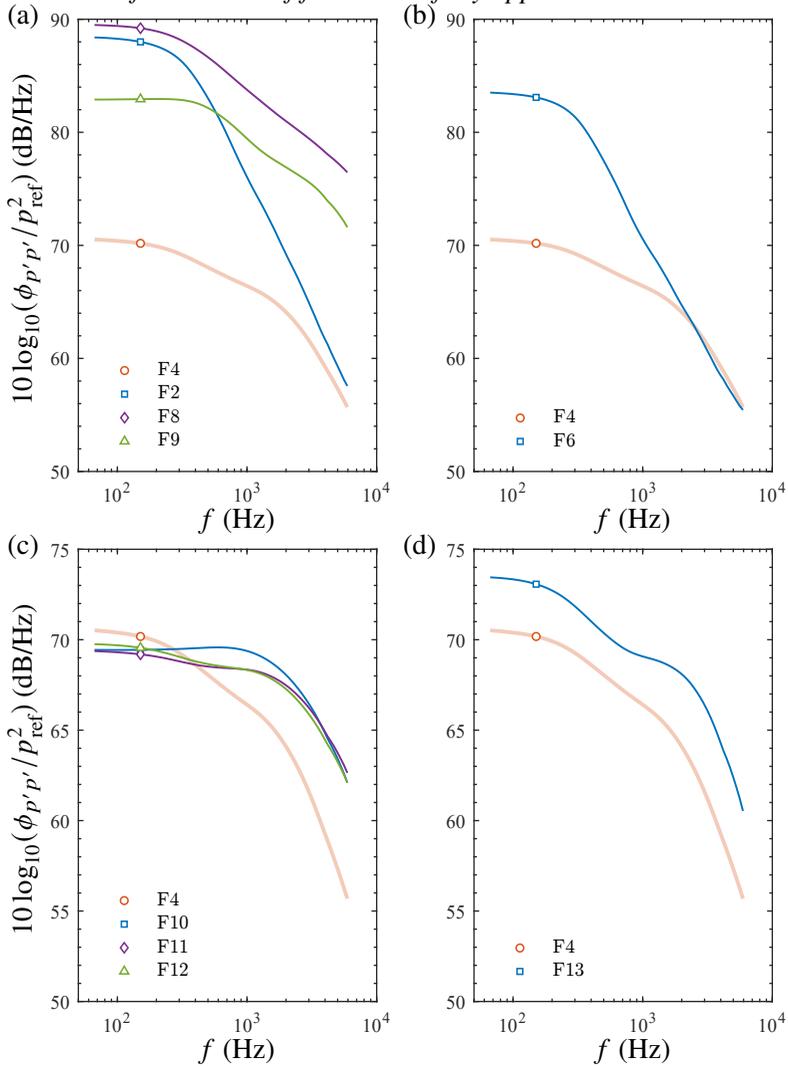

\paragraph{\textcolor{black}{Spatial evolution and fluid mechanics mechanisms}:} With the baseline condition established, we now examine the streamwise evolution of the pressure fluctuations along the model. Figure~\ref{fig:AFF8_hull_PSD} presents the streamwise evolution of pressure spectra, revealing complex history effects governing the boundary layer development. As shown in Figure~\ref{fig:AFF8_hull_PSD}(a), sensor F1 ($x/L=0.105$) exhibits significantly higher spectral levels than the mid-body baseline across the broadband range. Located shortly downstream of the trip wire ($x/L \approx 0.08$), this region is characterized by a  non-equilibrium turbulent boundary layer. The boundary layer here is extremely thin, placing the dominant turbulent source terms (in the buffer layer) physically very close to the wall sensor, thereby minimizing the distance-dependent decay and resulting in the observed energy overshoot.

\textcolor{black}{In the parallel mid-body region (F3, F4, F5), the spectral evolution reveals critical insights into the relaxation of the sail wake, as shown in Figure~\ref{fig:AFF8_hull_PSD}(b). A notable distinction is observed between the upstream and downstream sensors. 
Although F3 ($x/L=0.34$) and F4 ($x/L=0.50$) are both located within the ZPG parallel mid-body, their spectral characteristics differ significantly. 
F3 exhibits distinctly higher energy levels in the low-to-mid frequency range compared to F4. 
This amplification is physically attributed to the large-scale coherent vortex structures shed from the sail (located upstream at $x/L \approx 0.24$). 
These large-scale structures, convecting downstream in the outer layer of the boundary layer, imprint strong low-frequency pressure signatures on the wall at F3. In contrast, as the flow develops further downstream, the boundary layer recovers from this interference. 
The spectrum at F4 collapses perfectly with that of F5 ($x/L=0.71$). 
This spectral coincidence indicates that by $x/L=0.5$, the energetic influence of the sail wake has sufficiently decayed or diffused, and the turbulent boundary layer has reached a self-preserving equilibrium state. 
Consequently, F4 represents the true canonical turbulent boundary layer reference for this hull geometry, free from immediate appendage wake effects.}

\textcolor{black}{In the stern region (see Figure~\ref{fig:AFF8_hull_PSD}(c)), the spectrum at F7 ($x/L=0.956$) demonstrates a distinct ``crossover'' behavior relative to the baseline F4, driven by the geometric influence and the resulting adverse pressure gradient. 
In the low-frequency domain ($f < 700$ Hz), the energy at F7 is significantly amplified compared to the baseline. This enhancement is physically governed by the strong APG at the stern and the wake flow from the stern appendages, which amplify the inviscid instability of the outer boundary layer. These mechanisms promote the growth of energetic large-scale coherent structures, resulting in the observed spectral lift at low frequencies. Conversely, in the high-frequency domain ($f > 700$ Hz), the spectral trend reverses, with the level at F7 dropping well below that of F4. This attenuation arises from the rapid thickening of the boundary layer ($\delta$) at the tapered stern. Since high-frequency fluctuations scale inversely with the boundary layer thickness, the spatially expanding boundary layer effectively suppresses the small-scale turbulence intensity near the wall in this region.}

In terms of the appendage-induced spectra, the presence of appendages introduces significant spectral amplification, as shown in Figure~\ref{fig:AFF8_Combined_Appendage_PSD}. Sensors located near the sail-hull junction (F2, F8, F9) exhibit a massive amplification of low-frequency energy (Figure~\ref{fig:AFF8_Combined_Appendage_PSD}(a)). This corresponds to the dynamics of the horseshoe vortex system \citep{Jiang2025PoFSUBOFF} at the junction root. Similarly, the sensor at the stern-fin junction (F6, Figure~\ref{fig:AFF8_Combined_Appendage_PSD}(d)) shows increased levels compared to the baseline, reflecting the local flow separation and vortex shedding induced by the fins. Finally, Figure~\ref{fig:AFF8_Combined_Appendage_PSD}(c,d) presents the measurements taken directly on the appendage surfaces. The spectra on the sail (F10-F12) and the stern fin (F13) generally lie above the hull baseline F4. This enhancement is likely due to the thinner boundary layers developing on these lifting surfaces and the direct impingement of the freestream turbulence, resulting in more intense local pressure fluctuations.

\subsection{Impact of Appendages on Wall-Pressure Fluctuations}\label{Effects_of_appendages}
\begin{figure} 
    \centering
	\begin{minipage}[c]{1.0\linewidth}
    	\centering
    \begin{overpic}[width=4.0cm]{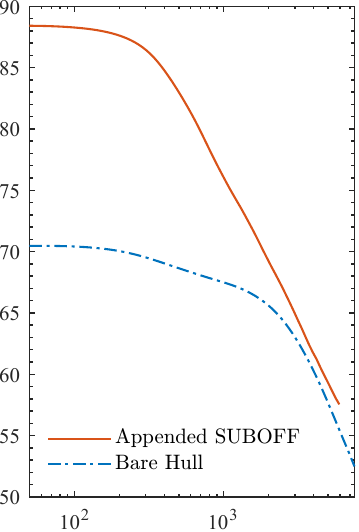}
        \put(-9,98){\color{black}{(a)}}
        \put(30,91){\color{black}\small{F2 (in APG-1)}}
        \put(-9,20){\small{\color{black}{\rotatebox{90}{$10 \log_{10}(\phi_{p^\prime p^\prime}/p^2_\text{ref})\, \text{(dB/Hz)}$}}}}
    \end{overpic}
    \hspace{0.4cm}
    \begin{overpic}[width=4.0cm]{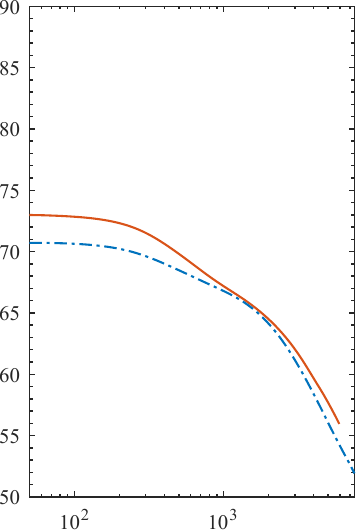}
        \put(-9,98){\color{black}{(b)}}
        \put(35,91){\color{black}\small{F3 (in ZPG)}}
    \end{overpic}
    \hspace{0.4cm}
    \begin{overpic}[width=4.0cm]{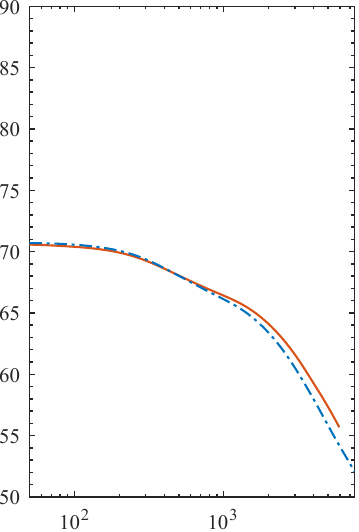}
        \put(-9,98){\color{black}{(c)}}
        \put(35,91){\color{black}\small{F4 (in ZPG)}}
    \end{overpic}
    \end{minipage}

    
	\begin{minipage}[c]{1.0\linewidth}
    	\centering
    \begin{overpic}[width=4.0cm]{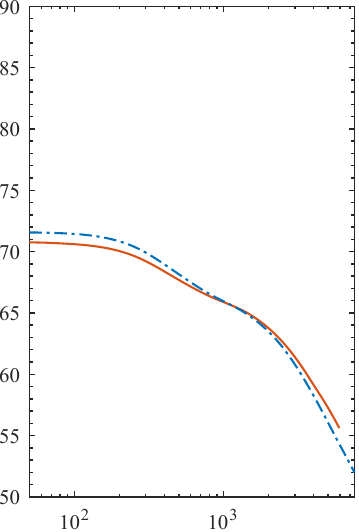}
        \put(-9,98){\color{black}{(d)}}
        \put(30,91){\color{black}\small{F5 (in FPG-1)}}
        \put(-9,20){\small{\color{black}{\rotatebox{90}{$10 \log_{10}(\phi_{p^\prime p^\prime}/p^2_\text{ref})\, \text{(dB/Hz)}$}}}}
        \put(29,-5){\small{\color{black}{$f$ (Hz)}}}
    \end{overpic}
    \hspace{0.4cm}
    \begin{overpic}[width=4.0cm]{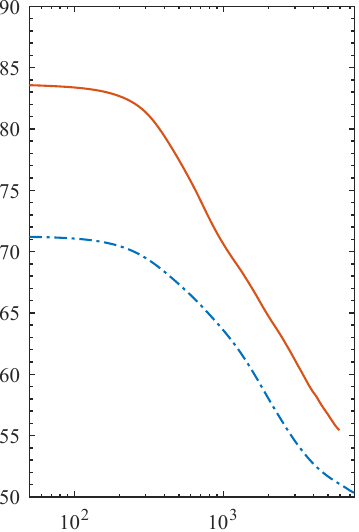}
        \put(-9,98){\color{black}{(e)}}
        \put(30,91){\color{black}\small{F6 (in APG-2)}}
        \put(29,-5){\small{\color{black}{$f$ (Hz)}}}
    \end{overpic}
    \hspace{0.4cm}
    \begin{overpic}[width=4.0cm]{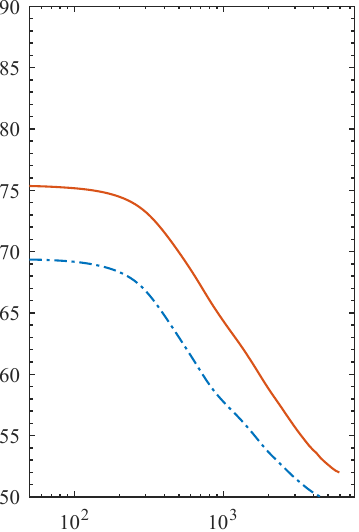}
        \put(-9,98){\color{black}{(f)}}
        \put(30,91){\color{black}\small{F7 (in FPG-2)}}
        \put(29,-5){\small{\color{black}{$f$ (Hz)}}}
    \end{overpic}
    \end{minipage}
    \vspace{0.5cm}
    \caption{Comparison of wall-pressure fluctuation power spectral densities between the fully appended SUBOFF model and the bare hull at $Re = 5.6 \times 10^6$. The comparison spans different flow regimes along the hull: (a) F2 in the forebody APG-1 region; (b) F3 and (c) F4 in the parallel mid-body ZPG region; (d) F5 in the aft-body FPG-1 region; and (e) F6 and (f) F7 in the stern APG-2/FPG-2 regions.}
    \label{fig:Comparison_AFF1_AFF8_20ms_F2-F7}
\end{figure}
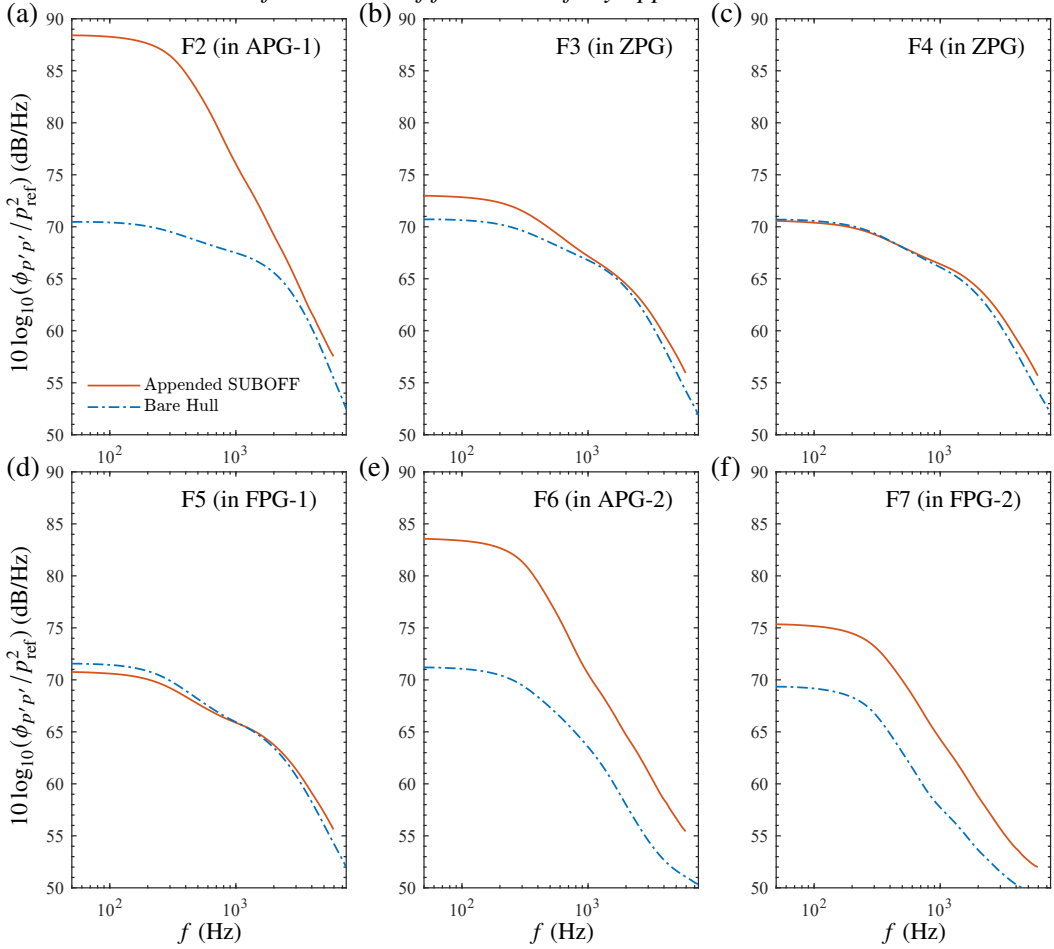
To quantify the hydrodynamic spectral amplification introduced by the appendages, a comparative analysis was conducted between the fully appended SUBOFF model and the bare hull baseline \citep{Jiang2025HullExperiment}. The presence of appendages significantly disturbs the boundary layer energy, particularly in the forebody and stern regions.

To facilitate a quantitative evaluation, the root-mean-square (RMS) fluctuating pressure, $p_{\text{rms}}$, is derived by integrating the one-sided power spectral density, $\phi_{p^\prime p^\prime}(f)$, over the effective frequency range ($f_{\min} = 50$ Hz to $f_{\max} = 6000$ Hz):

\begin{equation}
    p_{\text{rms}} = \sqrt{\int_{f_{\min}}^{f_{\max}} \phi_{p^\prime p^\prime}(f) \, df}.
    \label{eq:p_rms}
\end{equation}

This frequency band captures the dominant energy-containing eddies while excluding low-frequency facility noise and high-frequency sensor resonance. The fluctuating pressure coefficient, $C_p'$, is defined to allow for non-dimensional comparison:

\begin{equation}
    C_p' = \frac{p_{\text{rms}}}{q_{\infty}} = \frac{p_{\text{rms}}}{0.5 \rho U_{\infty}^2}.
    \label{eq:cp_prime}
\end{equation}

Furthermore, the percentage increase rate ($\eta$) relative to the bare hull is introduced to explicitly isolate the appendage contribution:

\begin{equation}
    \eta = \frac{C_{p,\text{app}}' - C_{p,\text{bare}}'}{C_{p,\text{bare}}'} \times 100\%,
    \label{eq:increase_rate}
\end{equation}

where the subscripts `app' and `bare' denote the appended and bare hull configurations, respectively.

\begin{figure} 
\centering
	\begin{minipage}[c]{1\linewidth}
     \centering
	\begin{overpic}[width=10cm]{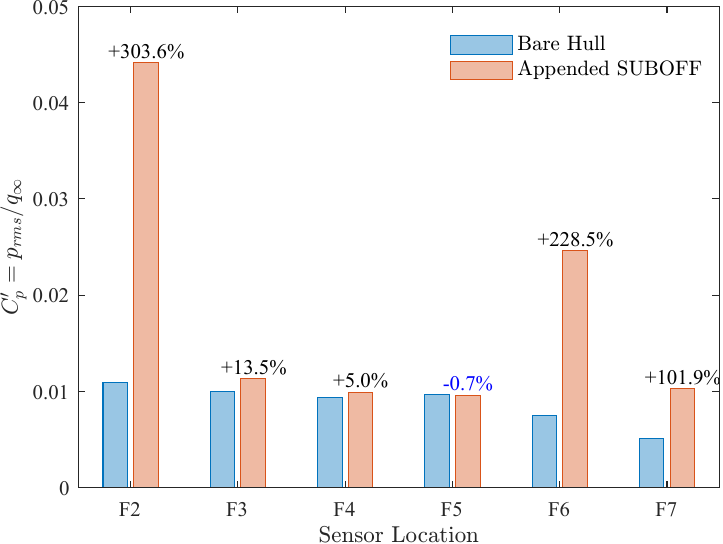}
	\end{overpic}
	\end{minipage}
    \caption{Comparison of the root-mean-square pressure coefficient ($C_p'$) between the fully appended SUBOFF model and the bare hull at $Re = 5.6 \times 10^6$.}
    \label{fig:AFF1_AFF8_Cp_RMS_BarChart}
\end{figure}

\textcolor{black}{The comparative analysis of the wall-pressure spectra (Figure~\ref{fig:Comparison_AFF1_AFF8_20ms_F2-F7}) and the integrated RMS statistics (Figure~\ref{fig:AFF1_AFF8_Cp_RMS_BarChart}) identifies the appendages as the dominant source of flow noise. The impact is highly localized and spatially evolving, initiating with the most intense interaction at the sail-hull junction. As illustrated in Figure~\ref{fig:AFF1_AFF8_Cp_RMS_BarChart}, the fluctuating pressure coefficient ($C_p'$) at the sail leading edge (Sensor F2) exhibits a massive increase of $\eta = +303.6\%$ relative to the bare hull. Spectrally, this amplification is concentrated in the low-frequency domain ($f < 1000$ Hz, see Figure~\ref{fig:Comparison_AFF1_AFF8_20ms_F2-F7}(a)). This phenomenon is physically driven by the adverse pressure gradient at the stagnation point, which triggers the formation of a highly unsteady horseshoe vortex system \citep{Jiang2025PoFSUBOFF}. These large-scale turbulent structures are the primary energetic contributors, inducing intense wall-pressure fluctuations at the junction root.}

\textcolor{black}{Downstream of the sail, the flow undergoes a gradual relaxation along the parallel mid-body. At Sensor F3 ($x/L=0.34$), located in the near-wake, a moderate increase of $\eta = +13.5\%$ persists. Figure~\ref{fig:Comparison_AFF1_AFF8_20ms_F2-F7}(b) indicates that this excess energy remains confined to low frequencies, corresponding to the convection of horseshoe vortex legs and sail wake eddies. However, as the flow evolves further downstream, the influence of the sail diminishes. By the time it reaches sensors F4 ($\eta = +5.0\%$) and F5 ($\eta = -0.7\%$), the RMS levels and spectral shapes (Figures~\ref{fig:Comparison_AFF1_AFF8_20ms_F2-F7}(c-d)) virtually collapse with the bare hull data. This recovery confirms that the boundary layer returns to a canonical equilibrium state, validating the use of these downstream locations as Zero Pressure Gradient references.}

\textcolor{black}{Hydrodynamic complexity re-emerges at the stern due to the presence of the stern fins. Sensor F6, positioned at the fin leading edge, records a substantial fluctuation increase of $\eta = +228.5\%$. Similar to the sail region, this is attributed to junction flow dynamics, though compounded here by a thicker incoming boundary layer. Crucially, at the tail tip (Sensor F7), fluctuations remain significantly elevated ($\eta = +101.9\%$). The broadband spectral amplification shown in Figure~\ref{fig:Comparison_AFF1_AFF8_20ms_F2-F7}(f) suggests that wake structures from the fins do not dissipate but rather converge and impinge upon the tapering tailcone. This interaction creates an energetic mixing layer that dominates the acoustic signature of the stern. In summary, appendages do not merely introduce a uniform background noise; rather, they generate specific, high-amplitude, low-frequency flow structures, that scale with the appendage dimensions, identifying them as the primary targets for hydrodynamic silence design.}

\subsection{Effectiveness of Vortex Control Baffle on Wall-Pressure Fluctuations}\label{Effects_of_vortex_control_baffle}

\begin{figure}
	\centering
	\begin{minipage}[c]{1\linewidth}
    	\centering
		\begin{overpic}[width=13cm]{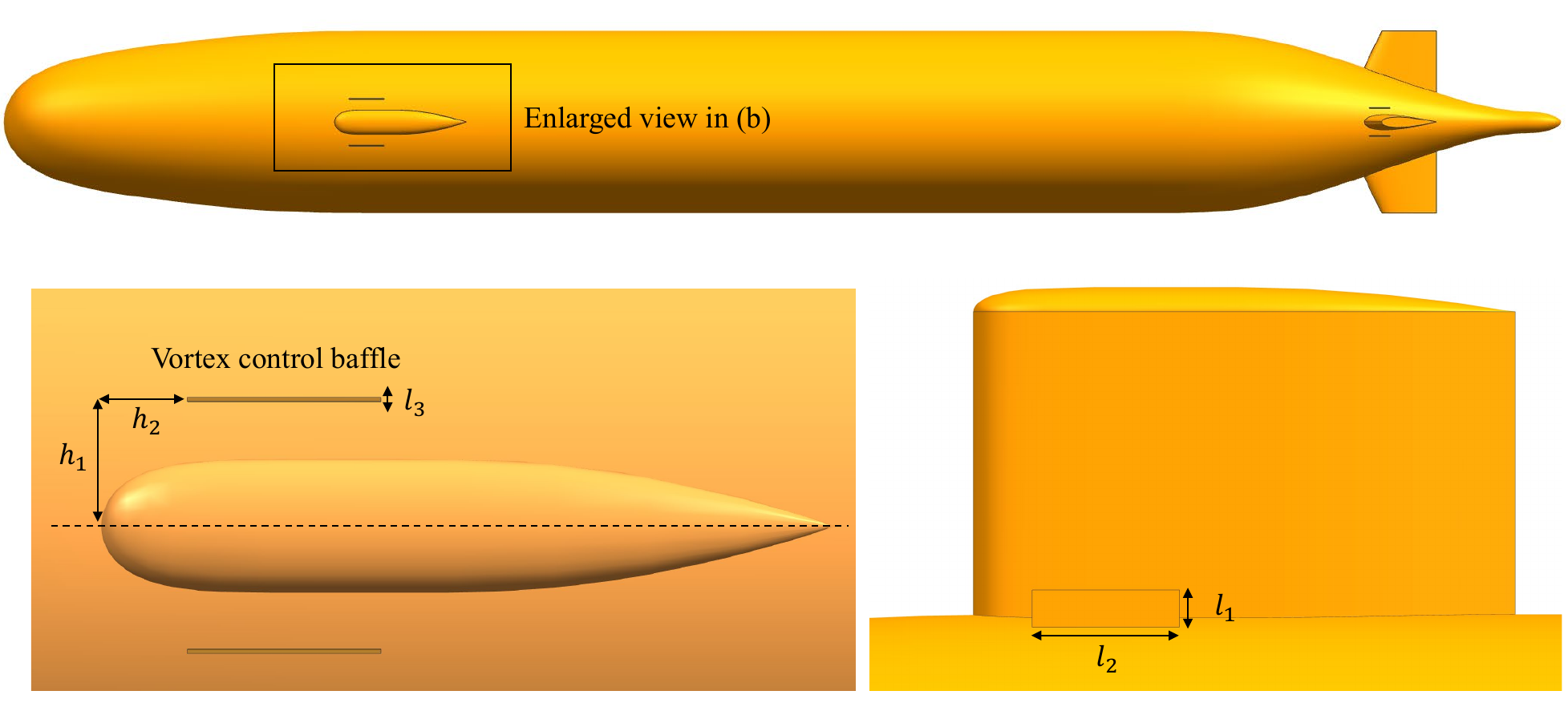}
			\put(-2,43){\color{black}{(a)}}
            \put(-2,27){\color{black}{(b)}}
            \put(58,27){\color{black}{(c)}}
		\end{overpic}
	\end{minipage}
    \caption{Schematic of the vortex control baffle configuration at the sail-hull junction: (a) Overall layout; (b) Top view and (c) side view detailing the baffle's dimensions.}
    \label{fig:Schematic_of_VCB}
\end{figure}

\begin{figure} 
    \centering
	\begin{minipage}[c]{1.0\linewidth}
    	\centering
    \begin{overpic}[width=4.0cm]{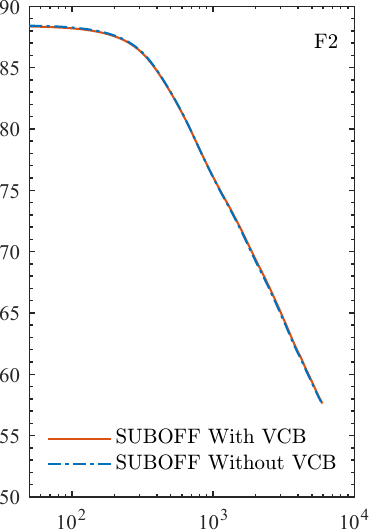}
        \put(-9,98){\color{black}{(a)}}
        \put(-9,20){\small{\color{black}{\rotatebox{90}{$10 \log_{10}(\phi_{p^\prime p^\prime}/p^2_\text{ref})\, \text{(dB/Hz)}$}}}}
    \end{overpic}
    \hspace{0.4cm}
    \begin{overpic}[width=4.0cm]{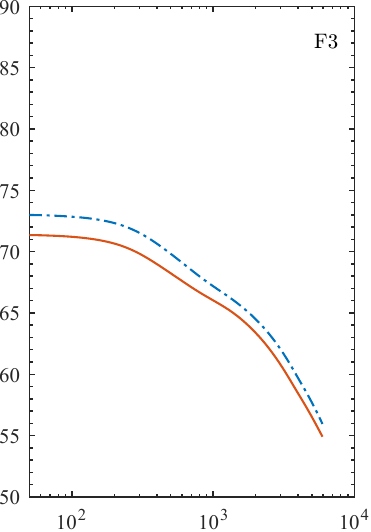}
        \put(-9,98){\color{black}{(b)}}
    \end{overpic}
    \hspace{0.4cm}
    \begin{overpic}[width=4.0cm]{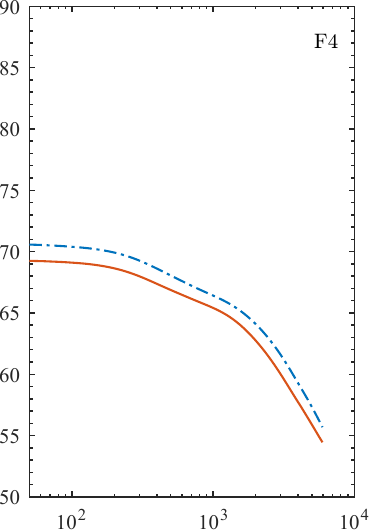}
        \put(-9,98){\color{black}{(c)}}
    \end{overpic}
    \end{minipage}

	\begin{minipage}[c]{1.0\linewidth}
    	\centering
    \begin{overpic}[width=4.0cm]{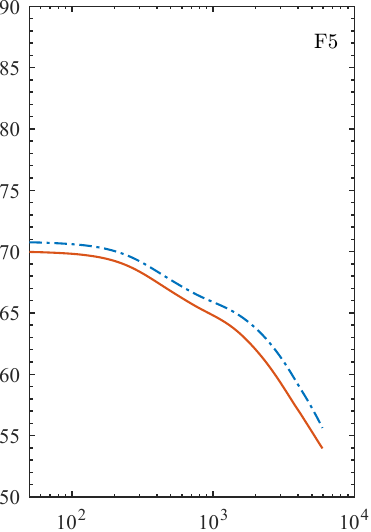}
        \put(-9,98){\color{black}{(d)}}
        \put(-9,20){\small{\color{black}{\rotatebox{90}{$10 \log_{10}(\phi_{p^\prime p^\prime}/p^2_\text{ref})\, \text{(dB/Hz)}$}}}}
        \put(29,-5){\small{\color{black}{$f$ (Hz)}}}
    \end{overpic}
    \hspace{0.4cm}
    \begin{overpic}[width=4.0cm]{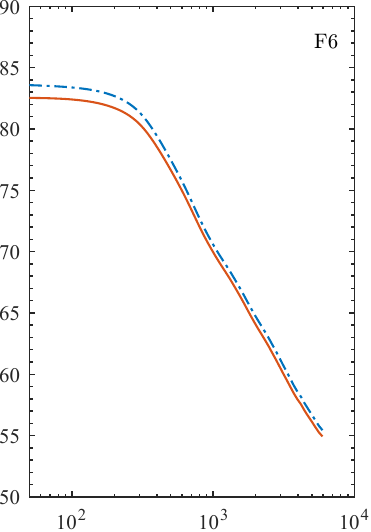}
        \put(-9,98){\color{black}{(e)}}
        \put(29,-5){\small{\color{black}{$f$ (Hz)}}}
    \end{overpic}
    \hspace{0.4cm}
    \begin{overpic}[width=4.0cm]{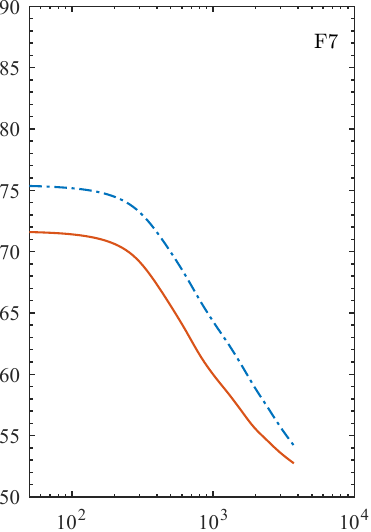}
        \put(-9,98){\color{black}{(f)}}
        \put(29,-5){\small{\color{black}{$f$ (Hz)}}}
    \end{overpic}
    \end{minipage}
    \vspace{0.5cm}
    \caption{Comparison of wall-pressure power spectral densities on the appended SUBOFF model without and with VCB at $Re = 5.6 \times 10^6$. (a) F2 at the sail leading edge; (b) F3 and (c) F4 along the parallel mid-body; (d) F5 at the aft-body transition; and (e) F6 and (f) F7 in the stern region.}
    \label{fig:Comparison_AFF8_AFF8_VCB_20ms_F3-F7}
\end{figure}
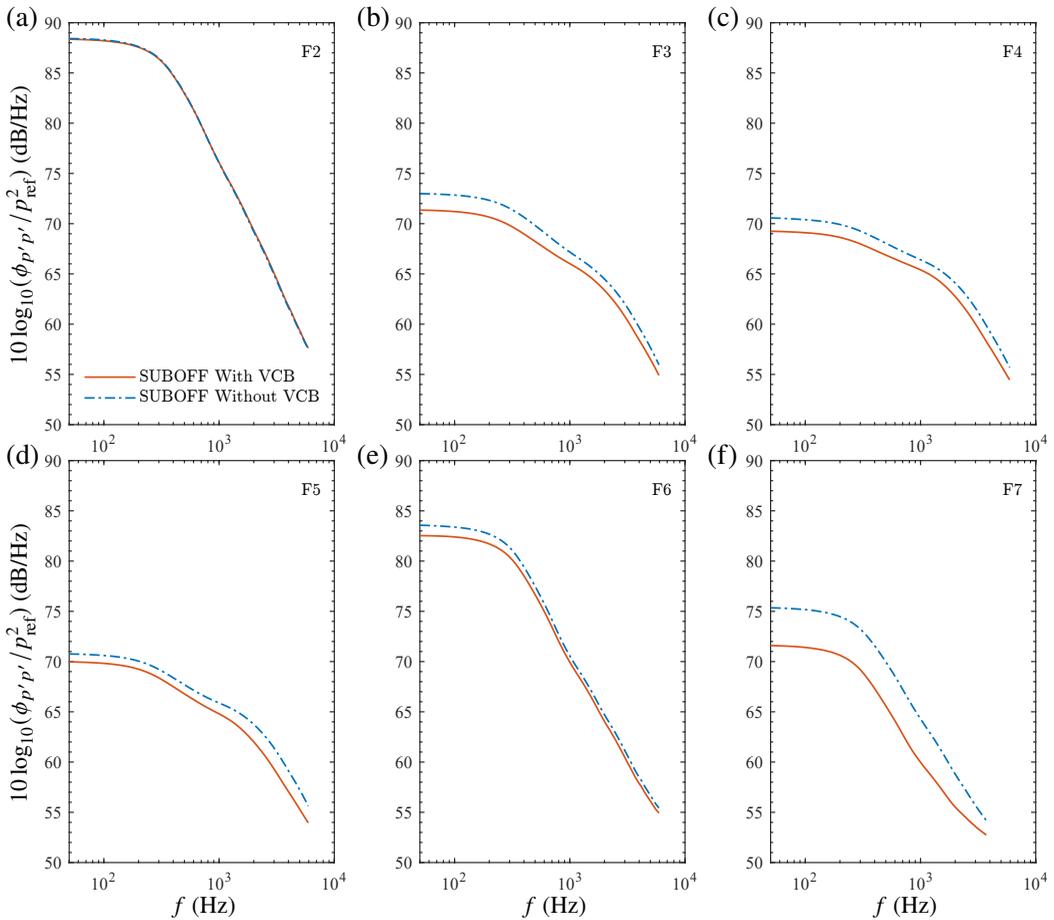

To suppress the appendage-induced flow noise identified in the previous section, a passive flow control strategy using vortex control baffles  was implemented. The working principle of the VCB is grounded in the disruption of the coherent horseshoe vortex structure. Previous hydrodynamic studies have established that to effectively suppress a horseshoe vortex, the control device must be precisely positioned at the vortex core. This placement disrupts the vortex formation process and induces a counter-rotating secondary flow that dissipates the high-intensity turbulent energy \citep{Wu2023VCBCFD}. Based on this principle, the geometric parameters of the VCB were optimized to match the local boundary layer characteristics and the trajectory of the vortex core \citep{liu2023control, liu2011numerical, Wu2023VCBCFD}. The configuration details are illustrated in Figure~\ref{fig:Schematic_of_VCB}. For the Sail VCB, the baffle features a streamwise length of $l_2 = 100\text{ mm}$, a height of $l_1 = 25\text{ mm}$, and a thickness of $l_3 = 2\text{ mm}$; the plates are oriented streamwise and positioned at a streamwise distance $h_2 = 40\text{ mm}$ from the leading edge of the sail, and a vertical distance $h_1 = 64\text{ mm}$ from the model center line. For the Stern VCB, the baffle features a streamwise length of $l_2 = 54\text{ mm}$, a height of $l_1 = 18\text{ mm}$, and a thickness of $l_3 = 2\text{ mm}$; these plates are positioned at a streamwise distance $h_2 = 15\text{ mm}$ from the leading edge of the stern appendage, and a vertical distance $h_1 = 38\text{ mm}$ from the model center line. These specific dimensions were selected to ensure the baffle is high enough to intercept the core of the horseshoe vortex but short enough to avoid generating excessive parasitic drag or strong wake-induced vortices. The baffles were aligned parallel to the incoming flow to minimize flow separation on the baffle surface itself.

To comprehensively evaluate the VCB in suppressing the pressure fluctuations, a comparative analysis was conducted between the baseline SUBOFF model and the VCB-installed configuration under identical flow conditions. The effectiveness of the flow control is quantified from two perspectives: the global reduction in fluctuation energy and the spectral-dependent insertion loss. To strictly quantify the effectiveness of the VCB, the fluctuating pressure coefficient ($C_p'$) defined in \S \ref{Effects_of_appendages} is utilized. While \S \ref{Effects_of_appendages} focused on the energy amplification caused by appendages, this section focuses on the reduction ratio ($\eta$), which is defined as:

\begin{equation}
\eta = \frac{C_{p,\text{base}}' - C_{p,\text{VCB}}'}{C_{p,\text{base}}'} \times 100\%
\end{equation}

where subscripts `base' and `VCB' denote the appended baseline and the controlled case, respectively. A positive $\eta$ indicates noise suppression. 

While $C_p'$ and $\eta$ provide a general view of energy reduction, they do not reveal the frequency-dependent mechanism of the flow control. To study how the VCB modulates turbulent structures of different scales (e.g., suppressing large-scale horseshoe vortices vs. affecting small-scale shear layer turbulence), the spectral Insertion Loss (IL) is introduced. The IL is defined as the difference in Sound Pressure Levels (SPL) between the baseline and the controlled cases at each frequency:

\begin{equation}
    \text{IL}(f) = \text{SPL}_{\text{base}}(f) - \text{SPL}_{\text{VCB}}(f) = 10 \log_{10}\left(\frac{\phi_{pp,\text{base}}(f)}{\phi_{pp,\text{VCB}}(f)}\right).
    \label{eq:insertion_loss}
\end{equation}

In this definition, a positive value ($\text{IL} > 0$) signifies effective noise reduction at a specific frequency $f$, whereas a negative value implies an increase in fluctuations, potentially due to additional flow disturbances introduced by the device itself. This method enables the identification of specific frequency bands where the VCB is most effective, thereby correlating the noise reduction with the suppression of characteristic flow structures such as the junction horseshoe vortex.

\begin{table}
    \centering
    \def~{\hphantom{0}}
    \begin{tabular}{l c c c c c}
        Sensor & $p_{\text{rms, base}}$ (Pa) & $p_{\text{rms, VCB}}$ (Pa) & $\eta$ (\%) & $\text{IL}_{\text{max}}$ (dB) & $f_{\text{peak}}$ (Hz) \\
        F1 & 5.85 & 5.91 & -1.05 & -- & -- \\
        F2 & 11.61 & 11.54 & 0.57 & -- & -- \\
        F3 & 2.85 & 2.43 & 14.71 & 1.71 & 286 \\
        F4 & 2.49 & 2.15 & 13.65 & 1.59 & 3315 \\
        F5 & 2.43 & 2.08 & 14.40 & 2.13 & 3514 \\
        F6 & 6.52 & 5.87 & 9.88 & 1.01 & 53 \\
        F7 & 2.65 & 1.72 & 35.16 & 4.48 & 728 \\
        F8 & 19.81 & 19.17 & 3.21 & -- & -- \\
        F9 & 11.91 & 10.81 & 9.18 & 1.83 & 5940 \\
        F10 & 3.62 & 3.73 & -3.04 & -- & -- \\
        F11 & 3.41 & 3.39 & 0.75 & -- & -- \\
        F12 & 3.34 & 3.32 & 0.54 & -- & -- \\
        F13 & 3.80 & 3.92 & -3.33 & -- & -- \\
    \end{tabular}
    \caption{Quantitative assessment of the vortex control baffle  effectiveness on wall-pressure fluctuations. The table compares the root-mean-square pressure ($p_{\text{rms}}$) integrated over $50$ -- $6000$\,Hz for the baseline and VCB configurations at $Re = 5.6 \times 10^6$. 
    Note: For upstream or stagnation sensors (e.g., F1, F2), the changes are negligible ($|\eta| < 5.0\%$), and thus no specific peak insertion loss frequency is assigned.}
    \label{tab:vcb_quantitative_results}
\end{table}

\textcolor{black}{While the spectral comparison of wall-pressure fluctuations with and without the VCB is provided in Figure~\ref{fig:Comparison_AFF8_AFF8_VCB_20ms_F3-F7}, the quantitative impact is systematically summarized in Table~\ref{tab:vcb_quantitative_results}. The analysis reveals a strong spatial dependence of the control effectiveness, confirming the VCB's role in modulating specific flow structures. In the upstream and stagnation regions (F1, F2), the VCB exhibits negligible influence, serving as a crucial validation of the experimental control. 
Sensors F1 (upstream) and F2 (junction leading edge) show minimal variations in RMS pressure, with reduction ratios ($\eta$) of $-1.05\%$ and $0.57\%$, respectively. 
The insertion loss spectra for these sensors fluctuate near zero across the entire frequency range (see Figure~\ref{fig:Comparison_AFF8_AFF8_VCB_20ms_F3-F7}(a)). 
This confirms that the VCB installation does not disturb the incoming flow or the mean stagnation pressure field, ensuring that the substantial noise reductions observed downstream are indeed attributed to the VCB's active flow control mechanisms rather than global experimental artifacts. Moving downstream to the junction wake and parallel mid-body (F3--F5, F9), a consistent and significant suppression emerges. 
At the rear of the junction (F9), a notable reduction of $9.18\%$ indicates that the VCB effectively weakens the formation of the primary horseshoe vortex legs as they wrap around the sail. 
This stabilizing effect propagates downstream, where sensors F3, F4, and F5 exhibit uniform reduction ratios ranging from $13.6\%$ to $14.7\%$. 
This sustained suppression confirms that the VCB successfully mitigates the shedding of large-scale coherent structures from the sail, thereby stabilizing the boundary layer development along the entire mid-section of the model. The control authority of the VCB culminates at the stern region (F6, F7), where the most dramatic effects are recorded. 
While Sensor F6 at the fin leading edge shows a moderate reduction ($9.9\%$), Sensor F7 at the tail tip exhibits a massive suppression of $35.16\%$. 
This result supports the physical hypothesis that, in the uncontrolled baseline case, the energetic horseshoe vortex legs persist downstream and impinge directly upon the tapered tail surface. 
The VCB effectively disrupts this vortex system at its source, eliminating this high-energy impingement mechanism and resulting in the substantial noise reduction observed at the tail.}

\begin{figure} 
    \centering
	\begin{minipage}[c]{1.0\linewidth}
    	\centering
    \begin{overpic}[width=4.2cm]{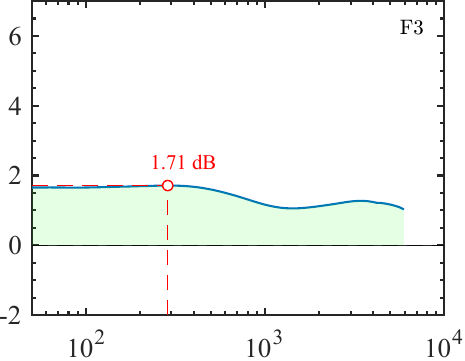}
        \put(-7,75){\color{black}{(a)}}
        \put(-7,14){\small{\color{black}{\rotatebox{90}{{Insertion Loss (dB)}}}}}
    \end{overpic}
    \hspace{0.1cm}
    \begin{overpic}[width=4.2cm]{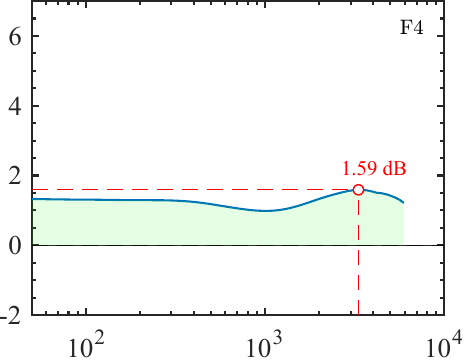}
        \put(-7,75){\color{black}{(b)}}
    \end{overpic}
    \hspace{0.1cm}
    \begin{overpic}[width=4.2cm]{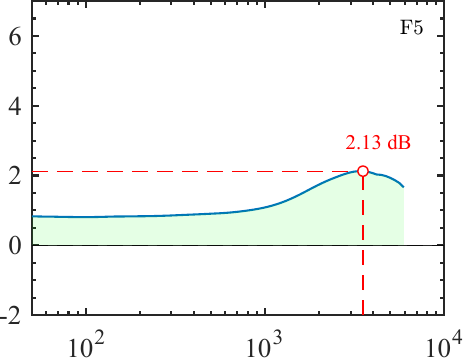}
        \put(-7,75){\color{black}{(c)}}
    \end{overpic}
    \vspace{0.2cm}
    \end{minipage}

	\begin{minipage}[c]{1.0\linewidth}
    	\centering
    \begin{overpic}[width=4.2cm]{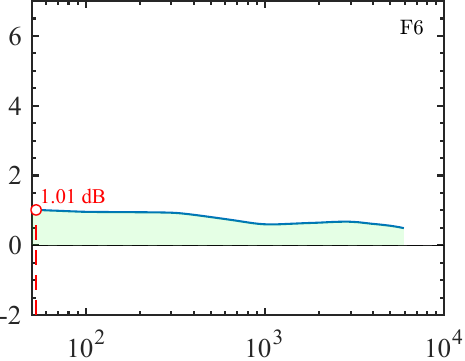}
        \put(-7,75){\color{black}{(d)}}
        \put(-7,14){\small{\color{black}{\rotatebox{90}{Insertion Loss (dB)}}}}
        \put(43,-7){\small{\color{black}{$f$ (Hz)}}}
    \end{overpic}
    \hspace{0.1cm}
    \begin{overpic}[width=4.2cm]{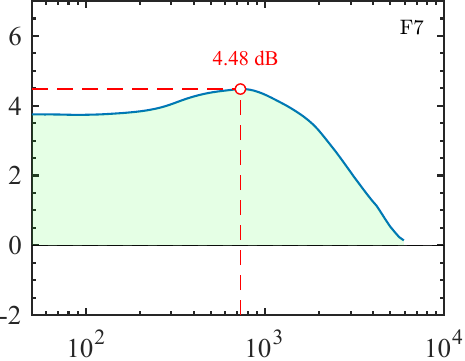}
        \put(-7,75){\color{black}{(e)}}
        \put(43,-7){\small{\color{black}{$f$ (Hz)}}}
    \end{overpic}
    \hspace{0.1cm}
    \begin{overpic}[width=4.2cm]{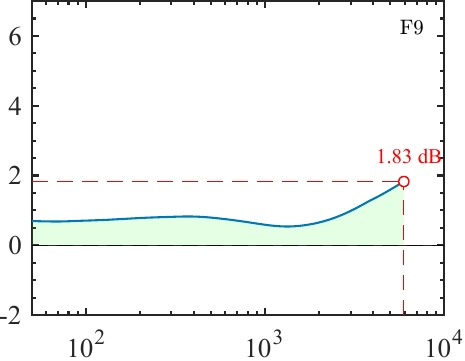}
        \put(-7,75){\color{black}{(f)}}
        \put(43,-7){\small{\color{black}{$f$ (Hz)}}}
    \end{overpic}
    \end{minipage}
    \vspace{0.5cm}
    \caption{Spectral insertion loss characteristics of the VCB at $Re = 5.6 \times 10^6$. Positive values ($\text{IL} > 0$, green shaded regions) indicate noise reduction. (a-c) Mid-body sensors F3, F4, and F5; (d) Stern APG sensor F6; (e) Stern tip sensor F7, and (f) Junction rear sensor F9. Red circles mark the maximum insertion loss ($\text{IL}_{\text{max}}$) frequencies.}
    \label{fig:IL_VCB}
\end{figure}
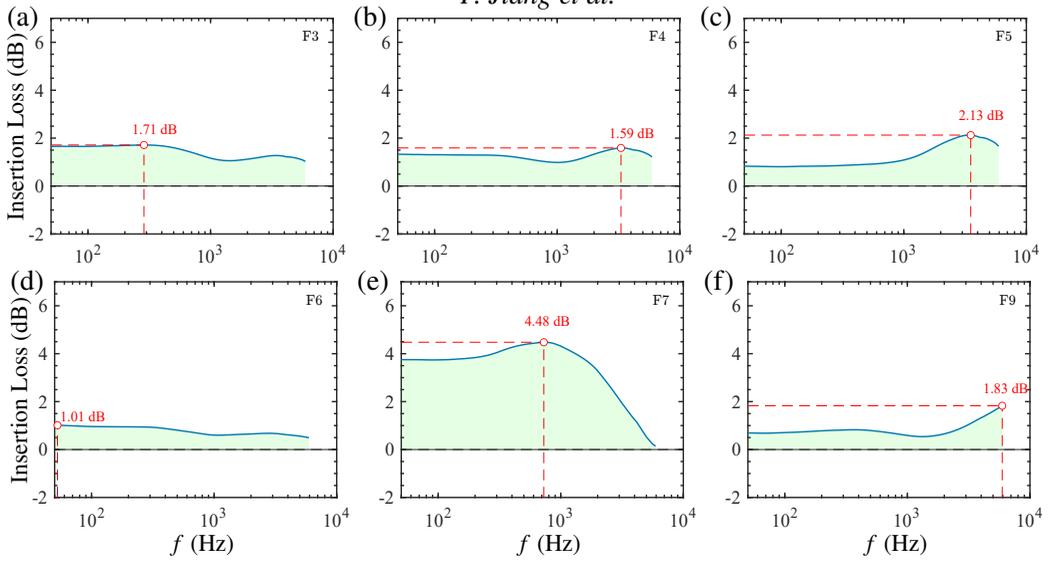

\textcolor{black}{To elucidate the physical mechanisms underlying the significant global energy reductions ($\eta$) identified in Table~\ref{tab:vcb_quantitative_results}, sensors exhibiting the most effective noise suppression (i.e., F9, F3--F6, and F7) were selected for detailed spectral analysis. 
By examining the insertion loss at these key locations, we can isolate the specific frequency bands where the VCB operates and correlate these changes with the modification of coherent turbulent structures. 
As illustrated in Figure~\ref{fig:IL_VCB}, the frequency-dependent performance reveals that the control mechanism is not uniform but varies distinctly across the flow regimes.
In the wake and parallel mid-body regions (sensors F3 through F6), the noise suppression manifests as a broadband phenomenon. 
The insertion loss remains positive across a wide bandwidth spanning from $100$\,Hz to $6$\,kHz, with peak reductions ($\text{IL}_{\text{max}} \approx 1.0$ -- $2.1$\,dB) occurring in the mid-frequency range ($f \approx 200$ -- $3500$\,Hz). 
This broad spectral signature corresponds to the energy-containing scales of the turbulent boundary layer. 
It indicates that the Sail VCB effectively attenuates the turbulent energy cascade within the wake, rather than targeting a single vortex shedding frequency, thereby stabilizing the global flow field.
In contrast, the spectral response at the tail tip (F7) reveals a more targeted control mechanism. 
The insertion loss spectrum is dominated by a distinct, high-amplitude peak of $4.48$\,dB at a specific frequency of $f \approx 728$\,Hz. 
This sharp elimination of energy confirms that the combined Sail and Fin VCBs function by physically breaking up coherent vortex structures. 
By disrupting these vortices at their source, the control strategy prevents their resonant impingement on the stern surface, providing a robust physical explanation for the massive total energy reduction observed in the statistical analysis.}

\subsection{Impact of Maneuvering Conditions on Wall-Pressure Fluctuations}\label{Effects_of_Maneuvering_Conditions}
The introduction of maneuvering attitudes, such as yaw and pitch, fundamentally breaks the axisymmetric equilibrium of the flow field established under straight-ahead conditions. This condition can induce complex three-dimensional (3D) effects, including crossflow instability, 3D flow separation, and asymmetric vortex shedding. This section focuses on maneuvers to study how these mechanisms redistribute wall pressure energy across the PSD.
\subsubsection{Effect of the yaw angle}\label{Yaw_Angle}
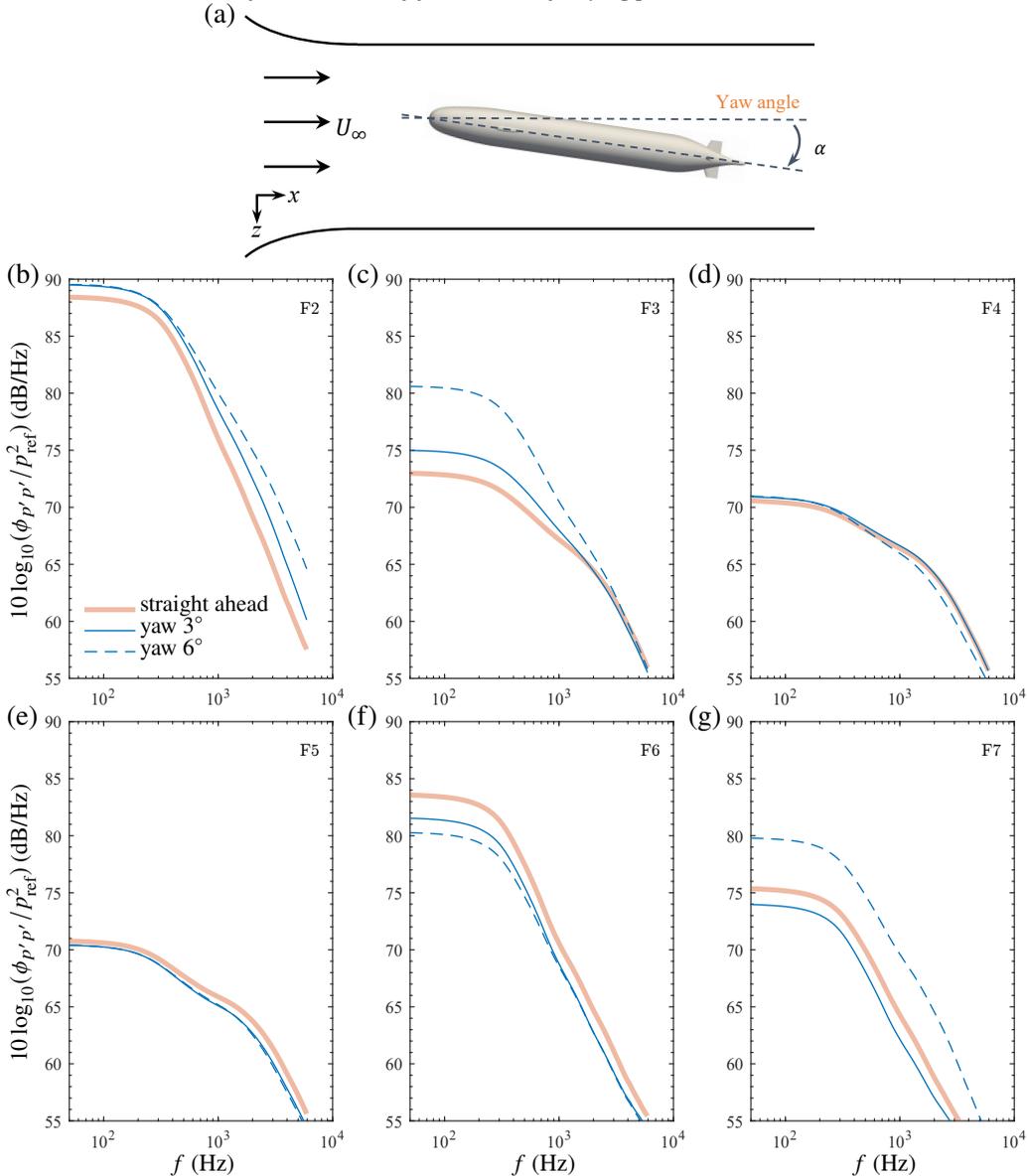
\begin{figure} 
    \centering
    \begin{minipage}[c]{1.0\linewidth}
    	\centering
    \begin{overpic}[width=8cm]{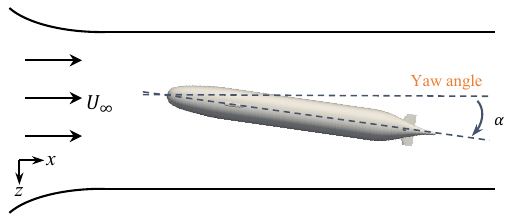}
        \put(-5,40){\color{black}{(a)}}
    \end{overpic}
    \end{minipage}
    
    \vspace{0.1cm}
	\begin{minipage}[c]{1.0\linewidth}
    	\centering
    \begin{overpic}[width=4.0cm]{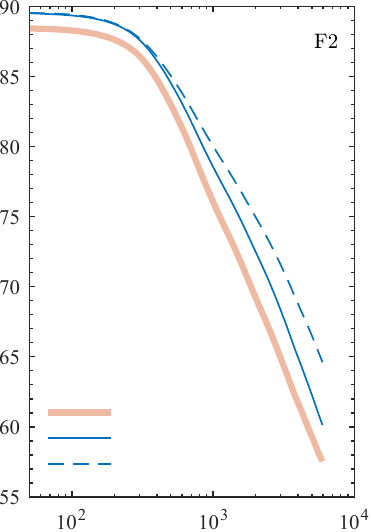}
        \put(-9,98){\color{black}{(b)}}
        \put(22,22){\footnotesize{\color{black}{straight ahead}}}
        \put(22,17){\footnotesize{\color{black}{yaw $3^\circ$}}}
        \put(22,12){\footnotesize{\color{black}{yaw $6^\circ$}}}
        \put(-9,20){\small{\color{black}{\rotatebox{90}{$10 \log_{10}(\phi_{p^\prime p^\prime}/p^2_\text{ref})\, \text{(dB/Hz)}$}}}}
    \end{overpic}
    \hspace{0.3cm}
    \begin{overpic}[width=4.0cm]{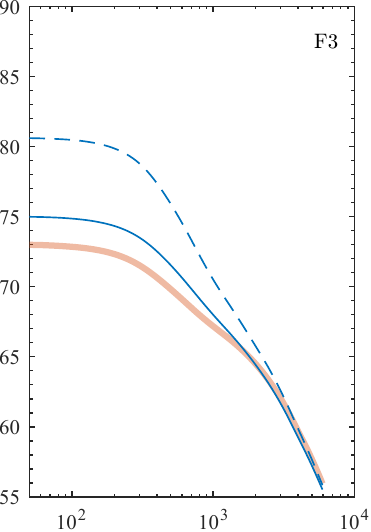}
        \put(-9,98){\color{black}{(c)}}
    \end{overpic}
    \hspace{0.3cm}
    \begin{overpic}[width=4.0cm]{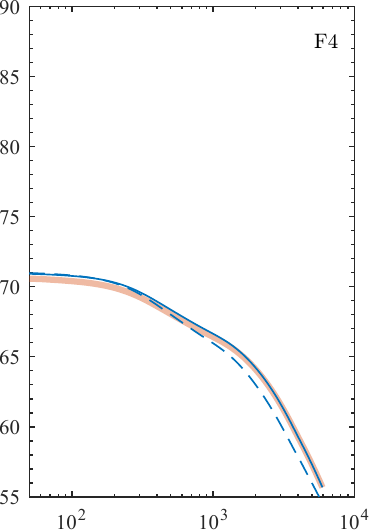}
        \put(-9,98){\color{black}{(d)}}
    \end{overpic}
    \end{minipage}

    \vspace{0.1cm}
    
	\begin{minipage}[c]{1.0\linewidth}
    	\centering
    \begin{overpic}[width=4.0cm]{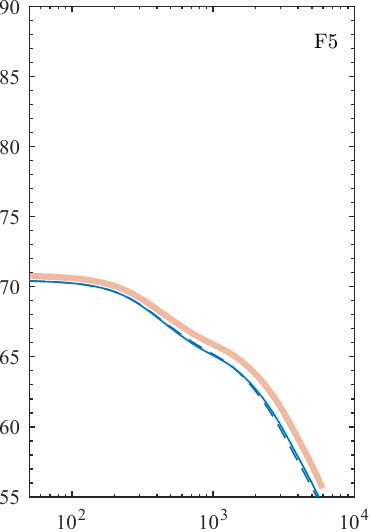}
        \put(-9,98){\color{black}{(e)}}
        \put(-9,20){\small{\color{black}{\rotatebox{90}{$10 \log_{10}(\phi_{p^\prime p^\prime}/p^2_\text{ref})\, \text{(dB/Hz)}$}}}}
        \put(29,-5){\small{\color{black}{$f$ (Hz)}}}
    \end{overpic}
    \hspace{0.3cm}
    \begin{overpic}[width=4.0cm]{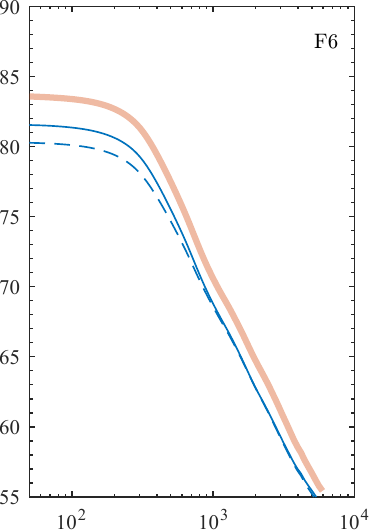}
        \put(-9,98){\color{black}{(f)}}
        \put(29,-5){\small{\color{black}{$f$ (Hz)}}}
    \end{overpic}
    \hspace{0.3cm}
    \begin{overpic}[width=4.0cm]{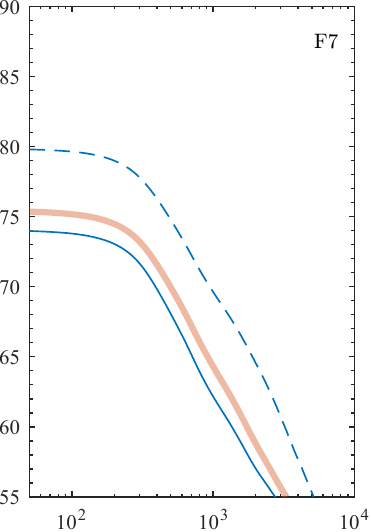}
        \put(-9,98){\color{black}{(g)}}
        \put(29,-5){\small{\color{black}{$f$ (Hz)}}}
    \end{overpic}
    \end{minipage}
    \vspace{0.2cm}
    \caption{Comparison of the PSD of pressure fluctuations under yaw angles with the straight ahead condition at $Re = 5.6 \times 10^6$. (a) Schematic of the yaw angle maneuver. Subfigures (b) F2, (c) F3, (d) F4, (e) F5, (f) F6, and (g) F7 illustrate sensors on the hull surface.}
    \label{fig:effect_yaw_angle_F2-F7}
\end{figure}

This section analyzes the response of wall-pressure fluctuations under horizontal turning (yaw) maneuvers. 
Figure \ref{fig:effect_yaw_angle_F2-F7} presents the power spectral density at yaw angles of $3^\circ$ and $6^\circ$, compared against the $0^\circ$ straight-ahead condition. 
The introduction of a yaw angle breaks the axisymmetric flow patterns, generating complex three-dimensional crossflow effects that manifest differently along the hull length.

In the forward hull and sail regions, the yaw maneuver induces a systematic amplification of pressure fluctuations. 
At the forward adverse pressure gradient region (F2), the PSD levels rise monotonically with the yaw angle (Figure \ref{fig:effect_yaw_angle_F2-F7}(b)), primarily in the mid-to-high frequency range ($300~\text{Hz} < f < 6~\text{kHz}$). 
This suggests that the crossflow component at the nose energizes the boundary layer, enhancing the production of small-scale turbulent structures. 
The spectral modification becomes even more pronounced immediately downstream of the sail (F3), where the presence of an angle of attack generates a robust turbulent wake (Figure \ref{fig:effect_yaw_angle_F2-F7}(c)). 
Here, the amplification is monotonic and substantial, reaching a maximum increase of approximately $8~\text{dB}$ at $6^\circ$ in the low-to-mid frequency range. 
This energy injection is attributed to the shedding of large-scale coherent structures and the unstable shear layer from the sail impinging directly on the hull surface.

In stark contrast to the disturbed sail region, the sensors along the parallel mid-body (F4 and F5) demonstrate a remarkable insensitivity to the yaw maneuver. 
As illustrated in Figures \ref{fig:effect_yaw_angle_F2-F7}(d) and (e), the spectra for the yawed conditions virtually collapse onto the straight-ahead baseline. 
This similarity to bare-hull behavior indicates that the distinct wake structures generated by the sail are either convected laterally by the crossflow or dissipate rapidly, leaving the flow field on the top meridian of the mid-body relatively undisturbed. 
This confirms that, under these moderate yaw angles, the strong influence of the appendage wake is spatially localized.

\begin{figure} 
    \centering
    \vspace{0.1cm}
	\begin{minipage}[c]{1\linewidth}
     \centering
	\begin{overpic}[width=6.5cm]{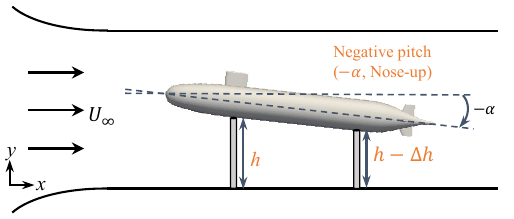}
        \put(-7,40){\color{black}{(a)}}
	\end{overpic}
    \hspace{0.2cm}
    \begin{overpic}[width=6.5cm]{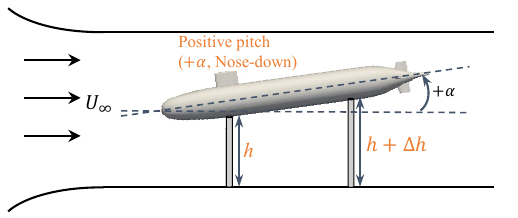}
        \put(-7,40){\color{black}{(b)}}
    \end{overpic}
    \vspace{0.3cm}
	\end{minipage}
	\begin{minipage}[c]{1.0\linewidth}
    	\centering
    \begin{overpic}[width=4.0cm]{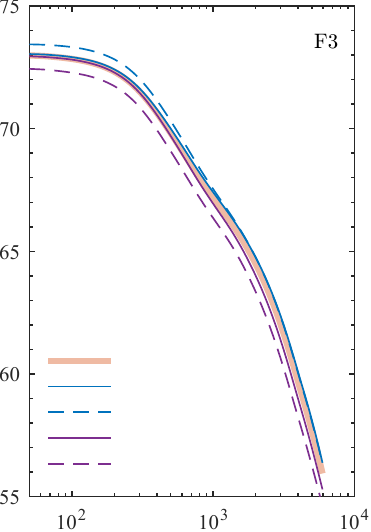}
        \put(-9,98){\color{black}{(c)}}
        \put(-9,20){\small{\color{black}{\rotatebox{90}{$10 \log_{10}(\phi_{p^\prime p^\prime}/p^2_\text{ref})\, \text{(dB/Hz)}$}}}}
        \put(29,-5){\small{\color{black}{$f$ (Hz)}}}
        \put(22,31){\footnotesize{\color{black}{straight ahead}}}
        \put(22,26){\footnotesize{\color{black}{pitch $-3^\circ$}}}
        \put(22,22){\footnotesize{\color{black}{pitch $-6^\circ$}}}
        \put(22,17){\footnotesize{\color{black}{pitch $+3^\circ$}}}
        \put(22,12){\footnotesize{\color{black}{pitch $+6^\circ$}}}
    \end{overpic}
    \hspace{0.3cm}
    \begin{overpic}[width=4.0cm]{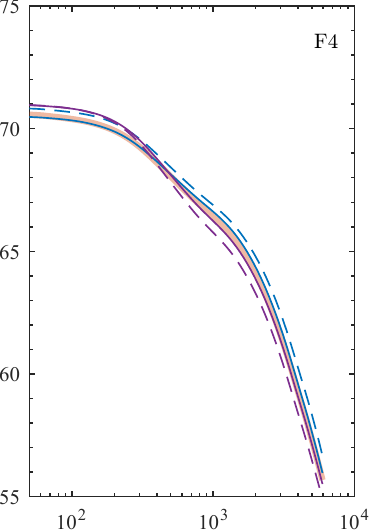}
        \put(-9,98){\color{black}{(d)}}
        \put(29,-5){\small{\color{black}{$f$ (Hz)}}}
    \end{overpic}
    \hspace{0.3cm}
    \begin{overpic}[width=4.0cm]{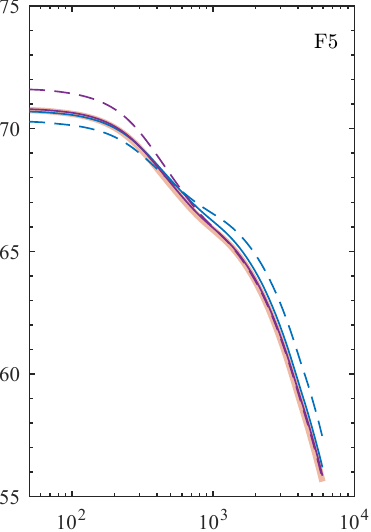}
        \put(-9,98){\color{black}{(e)}}
        \put(29,-5){\small{\color{black}{$f$ (Hz)}}}
    \end{overpic}
    \end{minipage}
    \vspace{0.5cm}
    \caption{Comparison of the PSD of pressure fluctuations under pitch angles with the straight ahead condition at $Re = 5.6 \times 10^6$. (a, b) Schematic of the yaw angle maneuver. Subfigures (c) F3, (d) F4, and (e) F5 illustrate sensors located in the middle body.}
    \label{fig:AFF8_pitch_F345}
\end{figure}
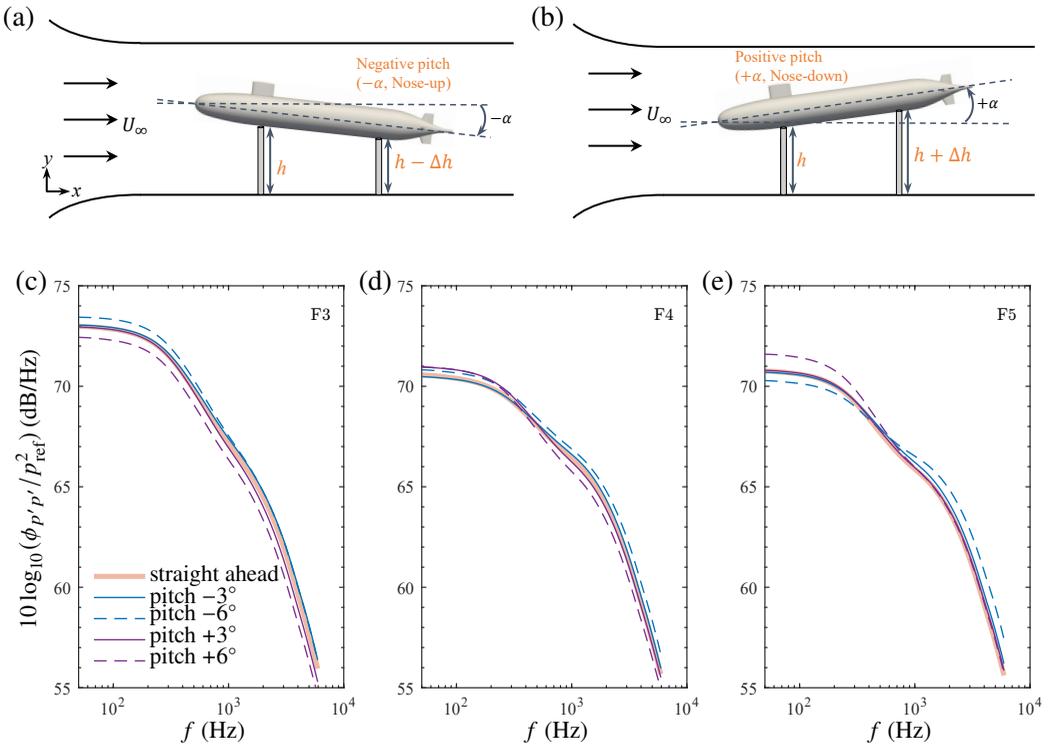

The response in the stern region, however, is complex and spatially dependent, governed by the interaction between developing crossflow vortices and the tapering hull. 
At the fin leading edge (F6), the yaw maneuver results in a distinct suppression of pressure fluctuations ($1 \sim 4~\text{dB}$), consistent with a ``shielding'' effect where stable crossflow vortices lift the primary shear layer away from the wall. 
Crucially, the flow behavior undergoes a non-monotonic transition at the extreme tail (F7). 
While a mild yaw angle of $3^\circ$ continues to show suppression, the trend reverses sharply at $6^\circ$, resulting in a broadband amplification of up to $5~\text{dB}$ (Figure \ref{fig:effect_yaw_angle_F2-F7}(g)). 
This non-monotonic variation signals a regime switch: at higher yaw angles, the strong 3D crossflow vortices likely detach or reorganize, leading to direct vortex impingement on the tail surface, thereby re-energizing the local pressure field.

\subsubsection{Effect of the pitch angle}\label{pitch_angle}

\begin{figure}
\centering
	\begin{minipage}[c]{1\linewidth}
     \centering
	\begin{overpic}[width=4.5cm]{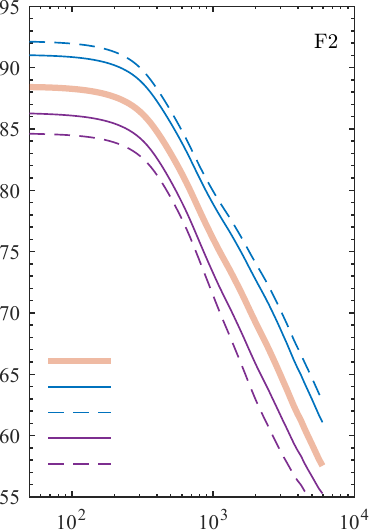}
        \put(-9,98){\color{black}{(a)}}
        \put(-9,20){{\color{black}{\rotatebox{90}{$10 \log_{10}(\phi_{p^\prime p^\prime}/p^2_\text{ref})\, \text{(dB/Hz)}$}}}}
        \put(22,31){\footnotesize{\color{black}{straight ahead}}}
        \put(22,26){\footnotesize{\color{black}{pitch $-3^\circ$}}}
        \put(22,22){\footnotesize{\color{black}{pitch $-6^\circ$}}}
        \put(22,17){\footnotesize{\color{black}{pitch $+3^\circ$}}}
        \put(22,12){\footnotesize{\color{black}{pitch $+6^\circ$}}}
	\end{overpic}
    \hspace{0.5cm}
    \begin{overpic}[width=4.5cm]{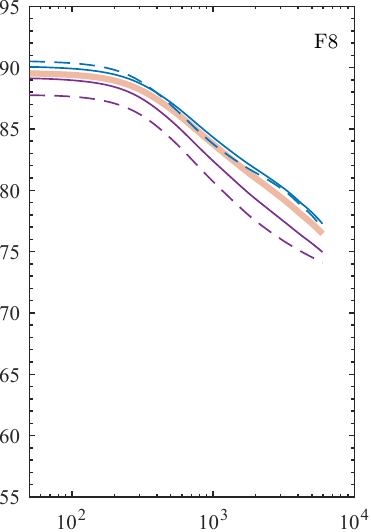}
        \put(-9,98){\color{black}{(b)}}
    \end{overpic}
	\end{minipage}

	\begin{minipage}[c]{1\linewidth}
     \centering
	\begin{overpic}[width=4.5cm]{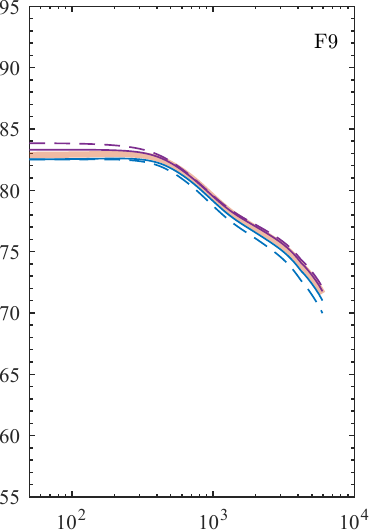}
        \put(-9,98){\color{black}{(c)}}
        \put(-9,20){{\color{black}{\rotatebox{90}{$10 \log_{10}(\phi_{p^\prime p^\prime}/p^2_\text{ref})\, \text{(dB/Hz)}$}}}}
        \put(30,-5){{\color{black}{$f$ (Hz)}}}
	\end{overpic}
    \hspace{0.5cm}
    \begin{overpic}[width=4.5cm]{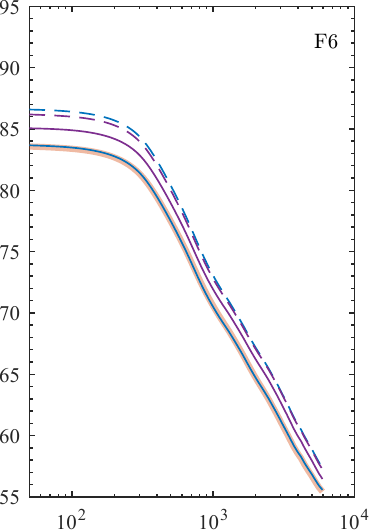}
        \put(-9,98){\color{black}{(d)}}
        \put(30,-5){{\color{black}{$f$ (Hz)}}}
    \end{overpic}
	\end{minipage}
    \vspace{0.5cm}
    \caption{Comparison of the PSD of pressure fluctuations under pitch angles with the straight ahead condition at $Re = 5.6 \times 10^6$. Subfigures (a) F2, (b) F8, (c) F9 show sensors located around the sail-hull junction, and (d) F6 illustrates sensor in front of the fin leading edge.}
    \label{fig:AFF8_pitch_F2689}
\end{figure}

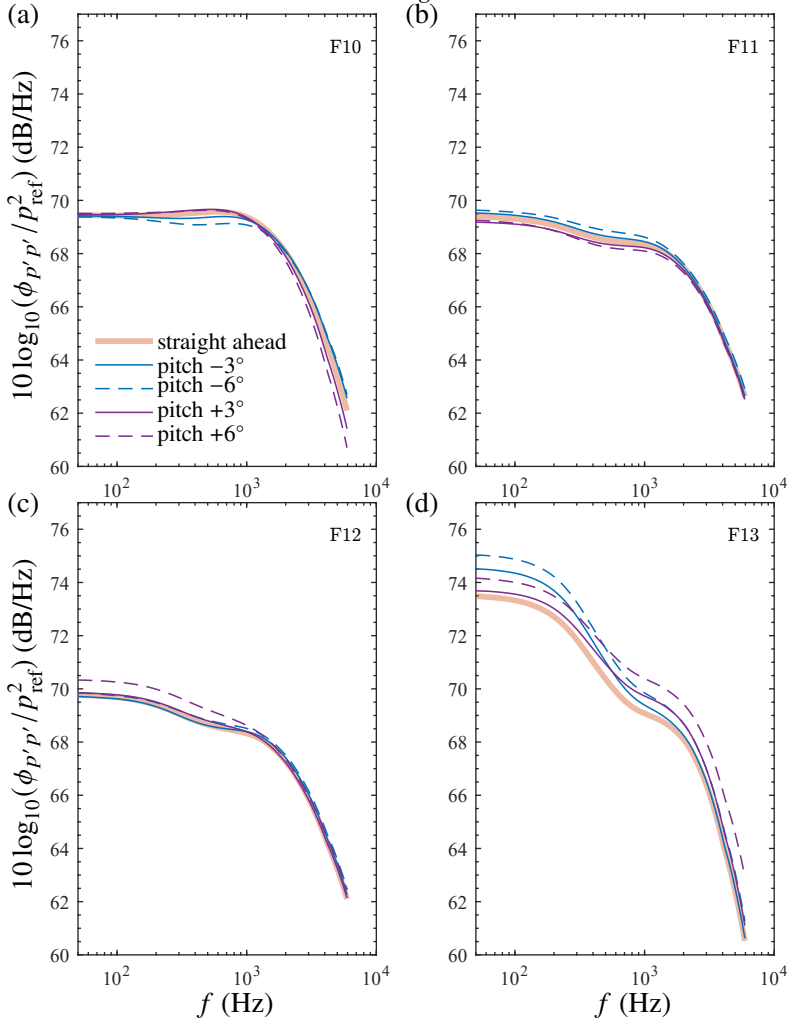
\begin{figure}
\centering
	\begin{minipage}[c]{1\linewidth}
     \centering
	\begin{overpic}[width=4.5cm]{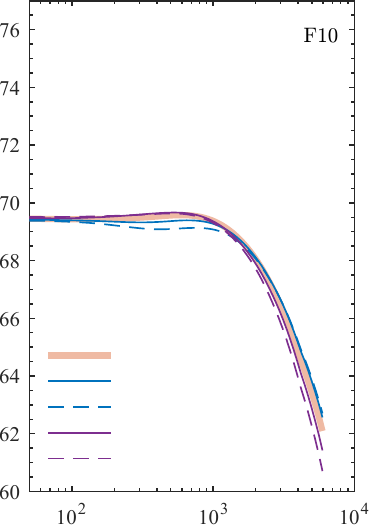}
        \put(-9,98){\color{black}{(a)}}
        \put(-9,20){{\color{black}{\rotatebox{90}{$10 \log_{10}(\phi_{p^\prime p^\prime}/p^2_\text{ref})\, \text{(dB/Hz)}$}}}}
        \put(22,31){\footnotesize{\color{black}{straight ahead}}}
        \put(22,26){\footnotesize{\color{black}{pitch $-3^\circ$}}}
        \put(22,22){\footnotesize{\color{black}{pitch $-6^\circ$}}}
        \put(22,17){\footnotesize{\color{black}{pitch $+3^\circ$}}}
        \put(22,12){\footnotesize{\color{black}{pitch $+6^\circ$}}}
	\end{overpic}
    \hspace{0.5cm}
    \begin{overpic}[width=4.5cm]{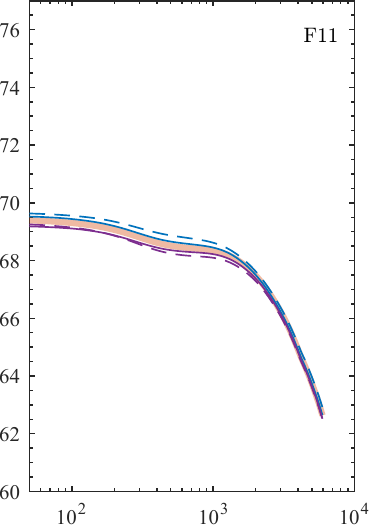}
        \put(-9,98){\color{black}{(b)}}
    \end{overpic}
	\end{minipage}

	\begin{minipage}[c]{1\linewidth}
     \centering
	\begin{overpic}[width=4.5cm]{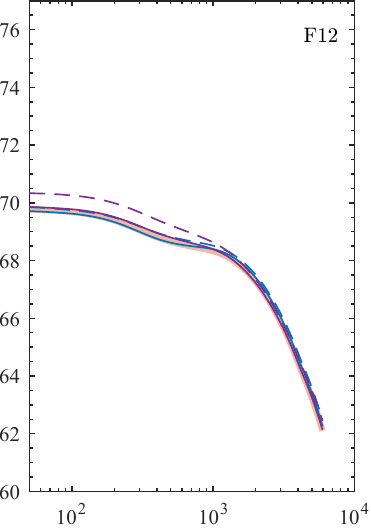}
        \put(-9,98){\color{black}{(c)}}
        \put(-9,20){{\color{black}{\rotatebox{90}{$10 \log_{10}(\phi_{p^\prime p^\prime}/p^2_\text{ref})\, \text{(dB/Hz)}$}}}}
        \put(30,-5){{\color{black}{$f$ (Hz)}}}
	\end{overpic}
    \hspace{0.5cm}
    \begin{overpic}[width=4.5cm]{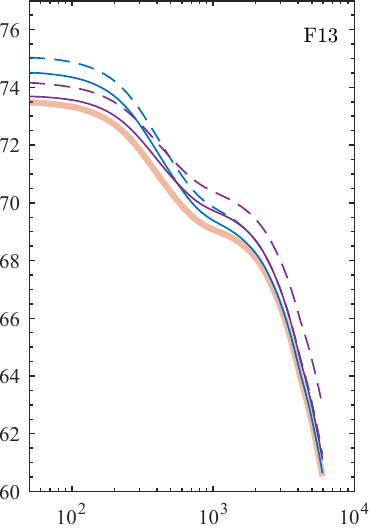}
        \put(-9,98){\color{black}{(d)}}
        \put(30,-5){{\color{black}{$f$ (Hz)}}}
    \end{overpic}
	\end{minipage}
    \vspace{0.5cm}
    \caption{Comparison of the PSD of pressure fluctuations under pitch angles with the straight ahead condition at $Re = 5.6 \times 10^6$. Subfigures (a) F10, (b) F11, (c) F12 show sensors located on the sail surface, and (d) F13 illustrates sensor on the fin surface.}
    \label{fig:AFF8_pitch_F10-13}
\end{figure}

The influence of the pitch maneuver on the wall-pressure fluctuations is analyzed by comparing the PSDs at various pitch angles ($ \pm3^\circ, \pm6^\circ$) against the straight-ahead baseline ($0^\circ$). Note that, in this study, a negative pitch angle corresponds to a nose-up attitude, while a positive pitch angle corresponds to a nose-down attitude, as shown in Figure \ref{fig:AFF8_pitch_F345}(a,b). The sensors are primarily located along the top meridian of the hull; consequently, during a negative pitch (nose-up) maneuver, the top surface acts as the leeward side, typically subjected to stronger adverse pressure gradients. Conversely, during a positive pitch (nose-down), the top surface acts as the windward side, experiencing favorable pressure gradients. The spectral responses are categorized by sensor location below.

As shown in Figure \ref{fig:AFF8_pitch_F345}, the sensors located in the parallel mid-body region demonstrate a relative insensitivity to pitch variations compared to the appendage regions. The PSD curves for both positive and negative pitch angles remain closely clustered around the straight-ahead baseline. Specifically, the negative pitch (nose-up) condition induces a uniform, slight amplification of the spectral levels (less than $1~\text{dB}$) across the frequency range. This is attributed to the thickening of the boundary layer on the leeward (top) side of the hull. Conversely, the positive pitch (nose-down) creates a windward effect, thinning the boundary layer and resulting in a marginal suppression of the pressure fluctuations. The minimal deviation suggests that the flow remains largely attached and equilibrium-like in the mid-body region, regardless of these moderate pitch attitudes.

The flow dynamics change drastically near the sail-hull junction, as illustrated in Figure \ref{fig:AFF8_pitch_F2689}(a-c). Sensor F2, located immediately upstream of the sail leading edge, exhibits the most pronounced sensitivity. A distinct, broadband spectral shift is observed, which varies monotonically with the pitch angle. The negative pitch (nose-up) significantly amplifies the fluctuations, with the $-6^\circ$ condition resulting in an increase of approximately $4~\text{dB}$ across the entire frequency spectrum. This indicates that the leeward orientation decelerates the incoming flow before it impinges on the sail, creating a highly energetic, turbulent stagnation region. Sensors F8 and F9, located alongside the sail, follow a similar trend to F2 but with reduced magnitude ($\sim 1~\text{dB}$ variation). This suggests that the intense turbulence generated at the sail leading edge (F2) gradually decays or diffuses as it convects downstream along the junction.

In the stern region, the flow response becomes non-monotonic with respect to the direction of pitch. At sensor F6 (Figure \ref{fig:AFF8_pitch_F2689}(d)), located upstream of the stern fin, both positive and negative pitch angles lead to an increase in PSD levels compared to the straight-ahead flight. Notably, this amplification is concentrated in the low-frequency range ($f < 300~\text{Hz}$). This behavior indicates that any deviation from the straight-ahead attitude disrupts the axisymmetric wake closure, inducing large-scale, low-frequency unsteady shedding or crossflow vortices that impact the fin root region.

The sensors mounted directly on the appendages show disparate behaviors depending on the appendage type, as shown in Figure \ref{fig:AFF8_pitch_F10-13}. (i) The pressure fluctuations on the sail surface are surprisingly stable. The variations are minimal ($<1~\text{dB}$), with negative pitch causing a slight rise in mid-to-low frequencies and positive pitch causing a slight drop. This implies that the flow attached to the sail sides remains relatively robust against the hull's pitch attitude. (ii) In contrast, the stern fin sensor (F13) is highly sensitive (Figure \ref{fig:AFF8_pitch_F10-13}(d)). Similar to F6, both pitch directions increase the fluctuation energy, but they affect different frequency bands. The negative pitch primarily amplifies the low-to-mid frequency content (up to $4~\text{dB}$), likely due to the ingestion of the thickened, turbulent hull wake (leeward side) into the fin surface. Conversely, the positive pitch primarily amplifies the mid-to-high frequency content ($\sim 4~\text{dB}$), which may be attributed to local flow separation or accelerated shear layer instabilities occurring on the fin surface itself under the windward hull orientation.

\section{Conclusions}\label{Conclusions}

This study establishes the first high-fidelity experimental database of wall-pressure fluctuations on the appended SUBOFF model, addressing critical gaps in high-Reynolds-number turbulence physics and maneuvering hydroacoustics. A systematic wind tunnel experiment was conducted on the fully appended DARPA SUBOFF model, with Reynolds numbers ranging from \(5.6 \times 10^6\) to \(1.4 \times 10^7\). The experimental setup included configurations both for baseline straight-ahead conditions and for complex maneuvering attitudes (yaw and pitch). A novel vortex control baffle  was evaluated for its impact on the wall-pressure fluctuations at the appendage-body junction. To ensure the accuracy of the experimental data, several critical calibration steps and Signal processing techniques, such as Wiener filtering, were applied to ensure that the pressure signals were clean and reliable.  To validate the fidelity of the aforementioned signal processing method, a benchmark experiment was conducted on a zero-pressure-gradient flat plate. The comprehensive agreement with a high-fidelity benchmark database across all spectral regimes provides a robust foundation for applying the same methodology to the more complex, three-dimensional flow fields of the appended SUBOFF model.

Based on the systematic analysis of the experimental data, the following conclusions are drawn:
\begin{enumerate}[label=(\roman*)]
\item The study reveals that within the high Reynolds number range investigated ($Re_L = 5.6 \times 10^6$ to $1.4 \times 10^7$), the wall-pressure fluctuation spectra exhibit remarkable self-similarity. The increase in spectral energy levels strictly follows the dynamic pressure scaling law ($p'_{rms} \propto U_\infty^2$) across the entire frequency band. This spectral consistency indicates that the turbulent boundary layer structures have reached a self-preserving equilibrium state in this regime. Key findings indicate that increasing the Reynolds number elevates the overall spectral levels while low-frequency content remains the dominant energetic contributor.

\item Appendages are quantified as the primary drivers of flow noise, introducing a spectral amplification compared to the bare hull. The study identifies two distinct amplification mechanisms: the formation of high-intensity horseshoe vortices at the appendage-hull junctions (causing a $\sim$300\% increase in fluctuations), and the downstream impingement of these coherent wake structures onto the tapered stern. This latter mechanism explains the re-amplification of noise at the tail, a critical insight for stern sonar array placement.

\item A major contribution of this work is the first successful experimental investigation of the VCB's effectiveness on a submarine geometry. The results demonstrate that by passively disrupting the coherence of the horseshoe vortex at its origin, the VCB achieves a global improvement in the downstream flow quality. This led to a substantial 35.16\% reduction in the root-mean-square pressure fluctuations at the stern and approximately 14\% along the parallel mid-body, proving that source-based flow control is a highly effective strategy for suppressing flow-induced noise.

\item The investigation into maneuvering states (yaw and pitch) reveals that the axisymmetric flow assumption is invalid during operation. The study uncovers complex, non-monotonic flow regimes driven by crossflow instability. Specifically, the stern region exhibits a transition from reducing noise due to flow separation at low angles to increased noise at higher angles. The results highlight the need to consider maneuvering conditions when predicting flow-induced noise, especially in operational scenarios where submarines undergo dynamic maneuvers.
\end{enumerate}

The findings of this study have several important implications for the design and operation of high-speed underwater vehicles. The experimental data provide a robust benchmark for validating numerical models of flow noise, particularly in complex geometries that include appendages and maneuvering conditions. Furthermore, the successful demonstration of the VCB's noise reduction capabilities presents a new direction for designing submarines with improved hydroacoustic stealth. This research also highlights the necessity of considering both steady and dynamic conditions when developing noise prediction models, as the latter can significantly influence the pressure field and, consequently, the hydroacoustic signature of the vehicle.

While this study provides important insights, there remain several areas for future study to advance our understanding and mitigation of flow-induced noise. (i) Future studies could focus on the optimization of VCB design parameters, exploring a broader range of VCB configurations to achieve a synergistic optimization of both noise reduction and wake uniformity. (ii) Another promising direction for future research is the measurement of wall non-dimensional velocity, which could provide a more detailed understanding of the relationship between wall-pressure fluctuations and the dynamics of the flow field. This would allow for the development of more accurate non-dimensional pressure spectra, which are crucial for improving the precision of noise prediction models. (iii) Finally, the incorporation of more complex numerical simulations, such as large eddy simulation (LES) or direct numerical simulation (DNS), in conjunction with experimental data, could provide deeper insights into the flow mechanisms that govern wall-pressure fluctuations. (iv) The non-monotonic pressure response under maneuvering conditions suggests that future hydroacoustic scaling laws must incorporate incidence-angle dependent correction factors to accurately predict operational noise.

\begin{bmhead}[Acknowledgements] 
This work was supported in part by National Natural Science Foundation of China (Grant No. 52522113), the Shanghai Pilot Program for Basic Research of Shanghai Jiao Tong University (No. 21TQ1400202), and the Fundamental Research Funds for the Central Universities.   
\end{bmhead}

\begin{bmhead}[Declaration of Interests] 
The authors report no conflict of interest.
\end{bmhead}

\begin{bmhead}[Data availability statement] 
The data that support the findings of this study are available from the corresponding author upon reasonable request.
\end{bmhead}

\begin{bmhead}[Author ORCIDs.] 
\\
\noindent Peng Jiang, \href{https://orcid.org/0000-0002-9072-8307}{https://orcid.org/0000-0002-9072-8307};\\
Bin Xie, \href{https://orcid.org/0000-0002-4218-2442}{https://orcid.org/0000-0002-4218-2442};\\
Shijun Liao, \href{https://orcid.org/0000-0002-2372-9502}{https://orcid.org/0000-0002-2372-9502}.
\end{bmhead}

\begin{appen}
\section{Detailed Signal Processing and Methodological Validation}\label{Signal_Processing}
This appendix provides the technical details of the multi-stage correction procedure summarized in \S \ref{Signal_Processing_main}, ensuring that the measured spectra accurately represent the physical wall-pressure fluctuations.

\subsection{Dynamic Response Correction for Pinhole Sensors}
\begin{figure}[h]
	\centering
	\begin{minipage}[c]{1\linewidth}
 	\centering
		\begin{overpic}[trim=1.7cm 1cm 6cm 0cm, clip, width=11cm]{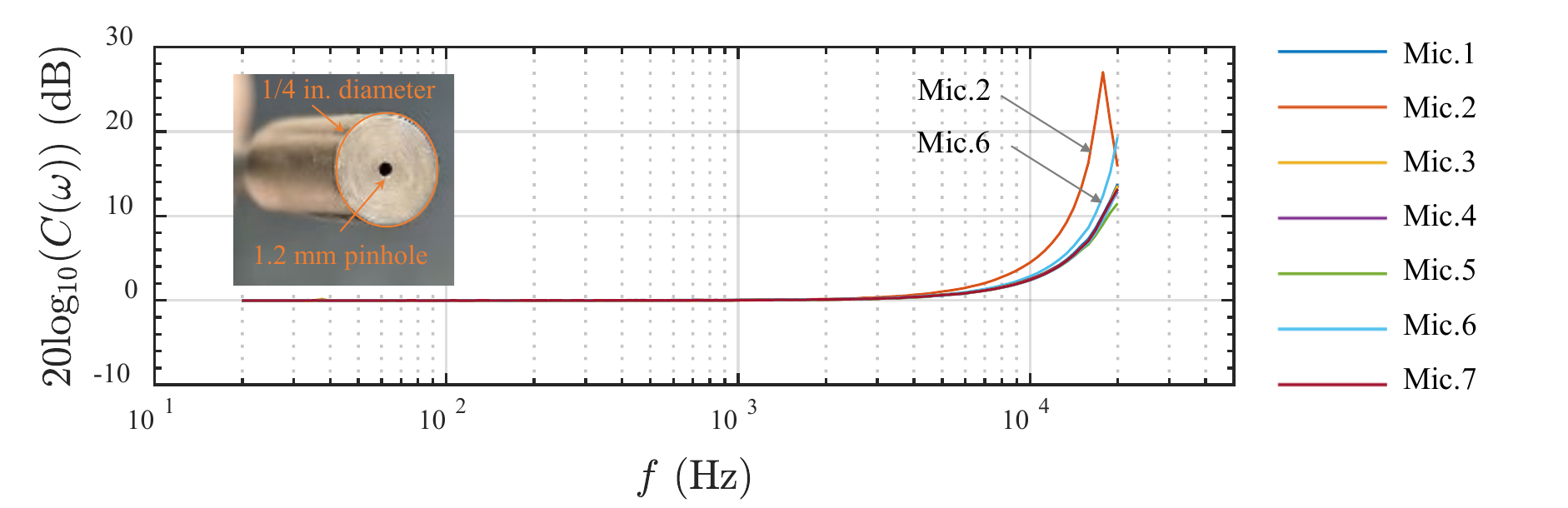}
		
        \put(-3,9){{\color{black}{\rotatebox{90}{$20 \log_{10}(C(f))\, \text{(dB)}$}}}}
        \put(50,-2){{\color{black}{$f$ (Hz)}}}
        \end{overpic}
	\end{minipage}
    \caption{The calibrated frequency response of the pinhole resonator for all the microphones.}
    \label{fig:frequency_response}
\end{figure}
Given the requirement to measure small-scale turbulent structures within the viscous sublayer, the spectral fidelity of the wall-pressure measurements is dependent upon the sensor's frequency response. The implementation of the pinhole-cap configuration \textcolor{black}{(1.2 mm diameter, see the inset of Figure~\ref{fig:frequency_response},} while essential for reducing spatial integration effects at high Reynolds numbers ($Re > 10^7$), introduces a Helmholtz resonance characteristic inherent to the cavity-neck system \citep{Tsuji2012Pressure}. This acoustic resonance results in a non-physical amplification of spectral energy in the high-frequency band, which has the potential to mask the true signals.

To correct the spectral distortion, a dynamic calibration was used. Unlike static sensitivity checks, this procedure characterizes the complex acoustic behaviors of the sensor system under broadband excitation. A transfer function was derived by cross-referencing the pinhole sensor's response against a laboratory-standard HBK 4192 microphone (flat response from 3.15\,Hz to 20\,kHz) in an anechoic environment based on the HBK 9721-A acoustic sensor calibration system. For dynamic correction, a continuous, frequency-dependent transfer function $C(f)$ was determined for each sensor via optimization. Following the Fourier-transformed rational model proposed by \citet{Joseph2017Pressure, Joseph2020JFMPlate}, this function is expressed as:

\begin{equation}
C(f) = \frac{b_2 f^2 + b_1 f + b_0}{a_2 f^2 + a_1 f + a_0}.
\label{eq:fitting model}
\end{equation}

\textcolor{black}{The coefficients in Eq.~\eqref{eq:fitting model} were adjusted to align the model with the calibration data, providing an accurate, continuous correction across the target frequency bandwidth. 
As illustrated in Figure~\ref{fig:frequency_response}, the fitted model shows excellent agreement with the experimental data for a representative sensor and captures the substantial resonant peak beyond 10 kHz. 
Consequently, the true hydrodynamic power spectral density, $\Phi_{pp}(f)$, is recovered via a spectral deconvolution process:}

\begin{equation}
 \Phi_{pp}(f) = \frac{\Phi_{measured}(f)}{|C(f)|^2}
 \label{eq:transfer_func}
\end{equation}

where $C(f)$ denotes the fitted transfer function.

\subsection{Background Noise Suppression via Wiener Filtering}
\begin{figure}
\centering
	\begin{minipage}[c]{1\linewidth}
     \centering
	\begin{overpic}[width=5.5cm]{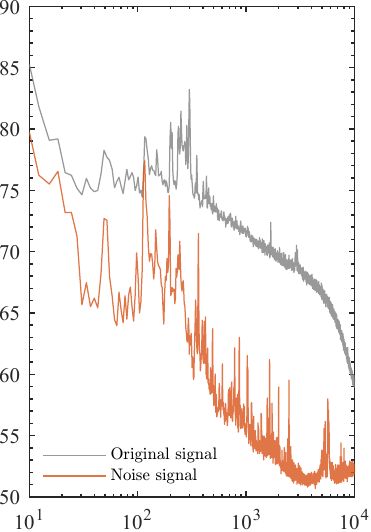}
        \put(-9,98){\color{black}{(a)}}
        \put(-9,20){{\color{black}{\rotatebox{90}{$10 \log_{10}(\phi_{p^\prime p^\prime}/p^2_\text{ref})\, \text{(dB/Hz)}$}}}}
        \put(30,-5){{\color{black}{$f$ (Hz)}}}
        \put(36,84){\includegraphics[trim=0cm 0cm 0cm 0cm, clip, scale=0.4]{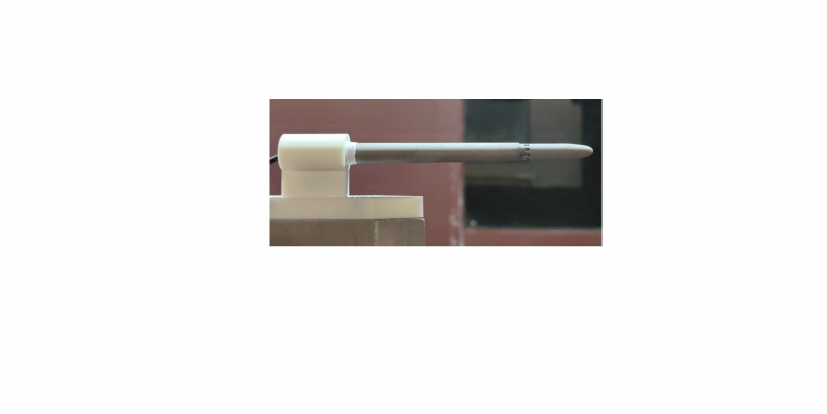}}
	\end{overpic}
    \hspace{0.5cm}
    \begin{overpic}[width=5.5cm]{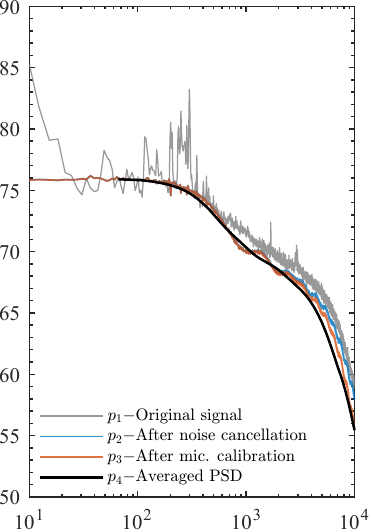}
        \put(-9,98){\color{black}{(b)}}
        \put(10,94){\color{black}{$U_\infty = 30$ m/s}}
        \put(10,88){\color{black}\footnotesize{dB re 20 $\mu$Pa}}
        \put(30,-5){{\color{black}{$f$ (Hz)}}}
    \end{overpic}
	\end{minipage}
    \vspace{0.5cm}
    \caption{Illustration of the wall-pressure signal correction procedure. (a) Comparison between the original wall-pressure signal and the facility background noise spectrum measured at $U_\infty = 30$ m/s, and the inset shows the device used to measure the facility background noise; (b) Step-by-step transformation of the wall-pressure PSD during the correction process: $p_1$ (Original signal PSD); $p_2$ (after noise cancellation); $p_3$ (after microphone dynamic response correction); and $p_4$ (final averaged and corrected PSD).}
    \label{fig:signal_correction_steps}
\end{figure}
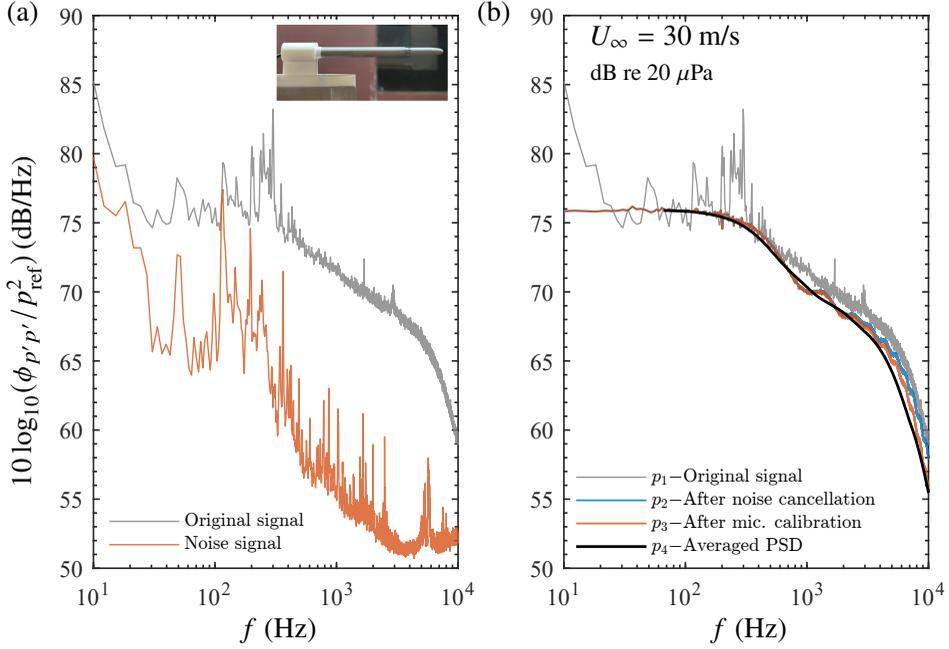

In closed-circuit wind tunnel facilities, acoustic contamination from the propulsion system and flow conditioning vanes propagates as plane waves, particularly dominating the low-frequency spectral range. 
To isolate the hydrodynamic pressure fluctuations from this facility background noise, a coherent noise suppression scheme based on the Wiener filtration theory was employed \citep{Baars2024JFMPlate}. 
A reference microphone (GRAS 46AE 1/2-inch freefield) equipped with a streamlined nosecone was positioned in the freestream, as shown in the inset of Figure~\ref{fig:signal_correction_steps}(a), to measure the global acoustic signature of the facility. 
Crucially, this sensor is insensitive to the local boundary layer turbulence on the model but maintains high coherence with the facility noise sources affecting the wall sensors. 
Subsequently, the optimal estimate of the uncontaminated wall-pressure spectrum is computed by:

\begin{equation}
\Phi_{clean}(f) = \Phi_{raw}(f) \left[ 1 - \gamma^2_{ref, wall}(f) \right],
\label{eq:wiener_filter}
\end{equation}

where $\gamma^2_{ref, wall}(f)$ represents the magnitude-squared coherence function between the wall-mounted sensor and the freestream reference. 
This approach is superior to direct spectral subtraction as it accounts for the phase relationship of the propagating acoustic waves. 
As depicted in Figure~\ref{fig:signal_correction_steps}(a), the original signal displays discrete tonal peaks at low frequencies, corresponding to the fan blade passing frequencies. 
The application of the Wiener filter yields the spectrum $p_2$, which successfully suppresses these acoustic tones by up to 15 dB without attenuating the broadband turbulent energy, ensuring an accurate representation of the large-scale structures.

In detail, Figure~\ref{fig:signal_correction_steps}(b) illustrates the processing steps of the wall-pressure signal through the multi-stage correction method. 
The original spectrum, denoted as $p_1$, is dominated by high-amplitude tonal peaks in the low-frequency range below 500 Hz, which corresponds to the harmonic frequencies of the wind tunnel propulsion fan. 
The noise-cancelled spectrum $p_2$ demonstrates the effectiveness of the Wiener filter. 
Note that the discrete tonal peaks are attenuated by approximately 10-15 dB, reducing the signal to the underlying broadband turbulent floor. 
Crucially, the slope of the spectrum in the convective ridge region remains unaltered, proving that the filter preserves flow energy. 
Applying the inverse transfer function to $p_2$ results in the corrected spectrum $p_3$. 
Here, the artificial resonance is completely flattened, restoring the physical decay characteristic of the viscous subrange. 
Finally, the spectral smoothing process yields the ultimate profile $p_4$. 
The PSD was computed using Welch's method with a Hanning window, 50\% overlap, and a block size of $2^{13}$ samples, yielding a frequency resolution of approximately 1.83 Hz.

\subsection{Methodological Validation against Canonical Benchmarks}

\begin{figure}[h]
\centering
    \begin{minipage}[c]{1.0\linewidth}
    	\centering
    \begin{overpic}[width=10cm]{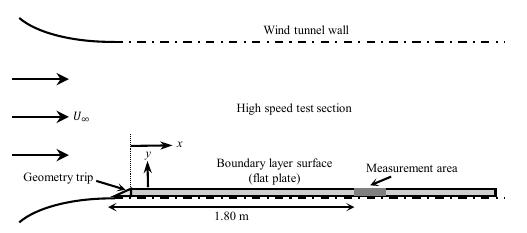}
        \put(-2,42){\color{black}{(a)}}
    \end{overpic}
        \vspace{0.5cm}
    \end{minipage}
	\begin{minipage}[c]{1\linewidth}
     \centering
	\begin{overpic}[width=5.0cm]{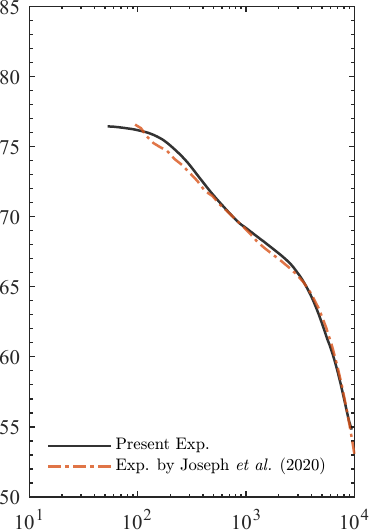}
        \put(-9,98){\color{black}{(b)}}
        \put(-9,20){{\color{black}{\rotatebox{90}{$10 \log_{10}(\phi_{p^\prime p^\prime}/p^2_\text{ref})\, \text{(dB/Hz)}$}}}}
        \put(30,-5){{\color{black}{$f$ (Hz)}}}
        \put(10,94){\color{black}{$U_\infty = 30$ m/s}}
        \put(10,88){\color{black}\footnotesize{dB re 20 $\mu$Pa}}
	   \end{overpic}
       \hspace{0.5cm}
    \begin{overpic}[width=5.0cm]{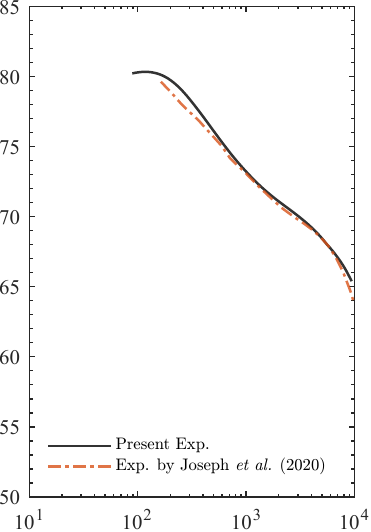}
        \put(-9,98){\color{black}{(c)}}
        \put(30,-5){{\color{black}{$f$ (Hz)}}}
        \put(10,94){\color{black}{$U_\infty = 40$ m/s}}
    \end{overpic}
	\end{minipage}
    \vspace{0.5cm}
    \caption{Validation of the wall-pressure signal correction methodology using a flat-plate turbulent boundary layer experiment. (a) Schematic of the test section configuration and coordinate system for the TBL experiment. (b) and (c) Comparison of the corrected wall-pressure power spectral density at $U_\infty = 30 $ m/s and $U_\infty = 40 $ m/s, respectively, with the benchmark data from  \citet{Joseph2020JFMPlate}.}
    \label{fig:validation_psd}
\end{figure}
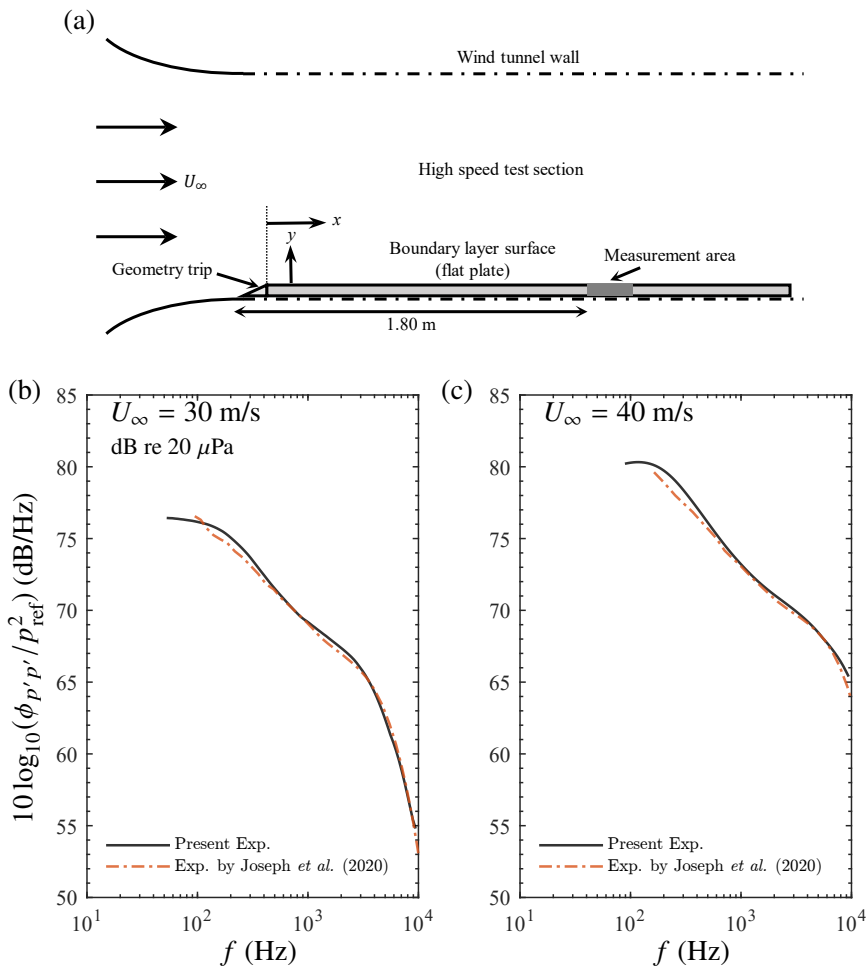

To validate the fidelity of the aforementioned signal processing method, specifically, the combined efficacy of the Wiener filter and pinhole correction, a benchmark experiment was conducted on a zero-pressure-gradient flat plate. This canonical flow configuration allows for a direct quantitative comparison with the high-fidelity database established by \citet{Joseph2020JFMPlate}, which is widely regarded as the standard for smooth-wall turbulent boundary layer pressure spectra. 

The validation tests were performed in the same facility at freestream velocities of $U_{\infty} = 30$ m/s and 40 m/s. Figure~\ref{fig:validation_psd} presents the comparison between the present corrected spectra (solid lines) and the benchmark data of \citet{Joseph2020JFMPlate} (dashed-dotted line). The agreement is significant: (i) In the logarithmic layer frequency band ($100 \text{ Hz} < f < 2000 \text{ Hz}$), the present data collapse remarkably well with the benchmark. This confirms that the sensor's spatial resolution is sufficient and the static calibration is accurate. (ii) At low frequencies, the excellent agreement confirms that the Wiener filter has successfully removed background noise, yielding a hydrodynamic signature comparable to that obtained in the anechoic facility of \citet{Joseph2020JFMPlate}. (iii) Most critically, the absence of any residual resonant peak in the high-frequency tail validates the accuracy of the pinhole transfer function $C(f)$. The spectral decay follows the expected viscous scaling, demonstrating that the deconvolution process has not introduced phase distortion or over-damping. This comprehensive agreement across all spectral regimes provides a robust foundation for applying the same methodology to the more complex, three-dimensional flow fields of the appended SUBOFF model.

\section{Estimation of Uncertainty}\label{uncertainty_estimation}
The determination of experimental uncertainty is paramount for validating the wall-pressure fluctuation data. The methodology for quantifying errors in this study adopts the approach detailed in our prior work on the bare-hull SUBOFF model \citep{Jiang2025HullExperiment}, which is rooted in established procedures \citep{Joseph2020JFMPlate}. This assessment integrates both Type A (random, statistical) and Type B (systematic, instrument-based) contributions, providing estimates at a 95\% confidence level using a coverage factor of $k=2$. The primary sources of uncertainty are analyzed below:

\begin{enumerate}[label=(\roman*)]
\item Transducer Specification: Systematic error (Type B) is introduced by the specified accuracy of the CYG1506G-P4LS12C2 sensors, which is $\pm 0.5$\% of the full scale ($\pm 2$\,kPa range). This results in an absolute uncertainty of $\pm 0.01$\,kPa. Assuming a uniform distribution, the standard uncertainty is calculated as $0.01 / \sqrt{3} \approx 0.0058$\,kPa, with the overall uncertainty reported using $k=2$.

\item Calibration Fitting Error: The dynamic response correction (Appendix~\ref{Signal_Processing}) relies on the fitting of a continuous transfer function. Analysis of the fitting residuals reveals a Root Mean Square Error (RMSE) of $0.0092$ across the $0$ -- 7.5\,kHz range. This systematic (Type B) error component is converted to a PSD uncertainty of approximately $\pm 0.2$\,dB at 95\% confidence, derived from $20 \log_{10}(1 + 2 \times 0.0092) \approx 0.16$\,dB (conservatively rounded).

\item Noise Cancellation Stability: The random uncertainty (Type A) associated with the Wiener filter (Appendix~\ref{Signal_Processing}) was quantified by examining the spectral output variability over 10 trials, changing the filter order around the nominal value ($m=16000 \pm 1000$). The resultant PSD variation ($\sigma$) near 1\,kHz was approximately $0.5$\,dB. The standard error for the noise cancellation is calculated as $u_{\text{noise}} = 0.5 / \sqrt{10} \approx 0.16$\,dB, which is then scaled by $k=2$.

\item Spectral Averaging and Processing: Given the sampling rate of 15\,kHz and a total record length of 90\,s, the processing utilizes approximately 329 segments with an 8192-point block size and 50\% overlap. This configuration results in a normalized random error of $\epsilon_r \approx 1/\sqrt{329} \approx 5.5$\% (or $\pm 0.24$\,dB) for the PSD estimates. While not negligible, this random component is subordinate to the systematic bias introduced by the Hanning window application, which is estimated at \textbf{$\pm 1.0$\,}dB for the current windowing parameters, consistent with findings in similar spectral analyses \citep{Joseph2020JFMPlate}.
\end{enumerate}

\begin{table}
	\centering
    \def~{\hphantom{0}}
	\begin{tabular}{lcc}
		Parameter & Uncertainty & Source \\
		Static pressure $p$ & $\pm 0.05$\,Pa & Transducer accuracy (i) \\
		Wall-pressure fluctuation $p'$ & $\pm 0.01$\,kPa & Sensor (i) \\
		RMS pressure $p_{\text{rms}}$ & $\pm 3.1$\% & PSD integration (ii--iv) \\
		PSD $\Phi(f)$ (low frequency) & $\pm 2.0$\,dB & Jitter, noise, calibration (ii--iv) \\
		PSD $\Phi(f)$ (high frequency) & $\pm 1.1$\,dB & Calibration, window bias (ii, iv) \\
	\end{tabular}
	\caption{Summary of uncertainties (95\% confidence). (Adapted from \citet{Jiang2025HullExperiment}).}
	\label{tab:uncertainty_summary}
\end{table}

The total uncertainty for the Root Mean Square (RMS) pressure ($p_{\text{rms}} = \sqrt{\langle p'^2 \rangle}$) is determined through standard error propagation. The relative uncertainty is expressed as $u_{p_{\text{rms}}} / p_{\text{rms}} \approx 0.5 \cdot u_{\Phi} / \Phi$. Converting the combined PSD uncertainties (dominated by the $\pm 1.0$\,dB window bias and $\pm 0.24$\,dB random error) into relative fractional errors, the integrated spectral uncertainty yields a final relative uncertainty of approximately $\pm 3.1$\% for $p_{\text{rms}}$.

The combined uncertainty for the Power Spectral Density (PSD), $u_{\Phi}$, is calculated by the root-sum-square method across the relevant Type A and Type B components ($\sqrt{u_{\text{random}}^2 + u_{\text{calib}}^2 + u_{\text{bias}}^2}$). This calculation yields an estimated uncertainty of $\pm 2.0$\,dB for the low-frequency range (accounting for higher noise cancellation variance) and approximately $\pm 1.1$\,dB for the high-frequency range ($\sqrt{1.0^2 + 0.2^2 + 0.24^2} \approx 1.05$\,dB). These final estimates are summarized in Table~\ref{tab:uncertainty_summary}.
\end{appen} 

\bibliographystyle{jfm}

\begin{thebibliography}{43}
\expandafter\ifx\csname natexlab\endcsname\relax\def\natexlab#1{#1}\fi
\def\au#1{#1} \def\ed#1{#1} \def\yr#1{#1}\def\at#1{#1}\def\jt#1{\textit{#1}} \def\bt#1{#1}\def\bvol#1{\textbf{#1}} \def\vol#1{#1} \def\pg#1{#1} \def\publ#1{#1}\def\arxiv#1{#1}\def\org#1{#1}\def\st#1{\textit{#1}}

\bibitem[Ashok {\em et~al.\/}(2015{\natexlab{{\em a\/}}})Ashok, Buren \& Smits]{ashok2015structure}
{\sc \au{Ashok, A.}, \au{Buren, T.~Van} \& \au{Smits, A.~J.}} \yr{2015{\natexlab{{\em a\/}}}}  \at{The structure of the wake generated by a submarine model in yaw}.  \jt{Exp. Fluids}  \bvol{56},  \pg{123}.

\bibitem[Ashok \& Smits(2013)]{ashok2013turbulent}
{\sc \au{Ashok, A.} \& \au{Smits, Alexander~J.}} \yr{2013} The turbulent wake of a submarine model in pitch and yaw.  \bt{In {\em 51st AIAA Aerospace Sciences Meeting including the New Horizons Forum and Aerospace Exposition\/}}. Grapevine (Dallas/Ft. Worth Region), Texas.

\bibitem[Ashok {\em et~al.\/}(2015{\natexlab{{\em b\/}}})Ashok, Van~Buren \& Smits]{ashok2015asymmetries}
{\sc \au{Ashok, A.}, \au{Van~Buren, T.} \& \au{Smits, A.~J.}} \yr{2015{\natexlab{{\em b\/}}}}  \at{Asymmetries in the wake of a submarine model in pitch}.  \jt{J. Fluid Mech.}  \bvol{774},  \pg{416--442}.

\bibitem[Baars {\em et~al.\/}(2024)Baars, Dacome \& Lee]{Baars2024JFMPlate}
{\sc \au{Baars, W.~J.}, \au{Dacome, G.} \& \au{Lee, M.}} \yr{2024}  \at{Reynolds-number scaling of wall-pressure-velocity correlations in wall-bounded turbulence}.  \jt{J. Fluid Mech.}  \bvol{981},  \pg{A15}.

\bibitem[Balantrapu {\em et~al.\/}(2023)Balantrapu, Alexander \& Devenport]{Balantrapu2023APG}
{\sc \au{Balantrapu, N.~A.}, \au{Alexander, W.~N.} \& \au{Devenport, W.}} \yr{2023}  \at{Wall-pressure fluctuations in an axisymmetric boundary layer under strong adverse pressure gradient}.  \jt{J. Fluid Mech.}  \bvol{960},  \pg{A28}.

\bibitem[Blake(2017)]{Blake2017Mechanics}
{\sc \au{Blake, W.~K.}} \yr{2017} {\em Mechanics of Flow-induced Sound and Vibration, Volume 2: Complex Flow--Structure Interactions\/}, ,  \vol{vol.~2}.  \publ{Academic Press}.

\bibitem[Ciappi \& Magionesi(2005)]{Ciappi2005characteristics}
{\sc \au{Ciappi, E.} \& \au{Magionesi, F.}} \yr{2005}  \at{Characteristics of the turbulent boundary layer pressure spectra for high-speed vessels}.  \jt{J. Fluids Struct.}  \bvol{21},  \pg{321--333}.

\bibitem[Corcos(1964)]{corcos1964structure}
{\sc \au{Corcos, G.~M.}} \yr{1964}  \at{The structure of the turbulent pressure field in boundary-layer flows}.  \jt{J. Fluid Mech.}  \bvol{18}~(3),  \pg{353--378}.

\bibitem[Fan {\em et~al.\/}(2025)Fan, Chen, Zhao \& Wan]{Fan2025ScalingJFM}
{\sc \au{Fan, G.~Q.}, \au{Chen, H.}, \au{Zhao, W.~W.} \& \au{Wan, D.~C.}} \yr{2025}  \at{Scaling laws and space-time characteristics of wall pressure fluctuations in an axisymmetric boundary layer with varying pressure gradient}.  \jt{J. Fluid Mech.}  \bvol{1016},  \pg{A28}.

\bibitem[Farabee \& Casarella(1991)]{farabee1991spectral}
{\sc \au{Farabee, T.~M.} \& \au{Casarella, M.~J.}} \yr{1991}  \at{Spectral features of wall pressure fluctuations beneath turbulent boundary layers}.  \jt{Phys. Fluids}  \bvol{3}~(10),  \pg{2410--2420}.

\bibitem[Gibeau \& Ghaemi(2021)]{Gibeau2021JFMPlate}
{\sc \au{Gibeau, B.} \& \au{Ghaemi, S.}} \yr{2021}  \at{Low- and mid-frequency wall-pressure sources in a turbulent boundary layer}.  \jt{J. Fluid Mech.}  \bvol{918},  \pg{A18}.

\bibitem[Goody(2004)]{Goody2004}
{\sc \au{Goody, M.}} \yr{2004}  \at{Empirical spectral model of surface pressure fluctuations}.  \jt{AIAA J.}  \bvol{42}~(9),  \pg{1788--1794}.

\bibitem[Goody(1999)]{Goody1999Experimental}
{\sc \au{Goody, M.~C.}} \yr{1999}  \at{An experimental investigation of pressure fluctuations in three-dimensional turbulent boundary layers}. PhD thesis, Virginia Polytechnic Institute and State University, Blacksburg, Virginia.

\bibitem[Gravante {\em et~al.\/}(1998)Gravante, Naguib, Wark \& Nagib]{Gravante1998dPlus}
{\sc \au{Gravante, S.~P.}, \au{Naguib, A.~M.}, \au{Wark, C.~E.} \& \au{Nagib, H.~M.}} \yr{1998}  \at{Characterization of the pressure fluctuations under a fully developed turbulent boundary layer}.  \jt{AIAA J.}  \bvol{36}~(10),  \pg{1775--1785}.

\bibitem[Groves {\em et~al.\/}(1989)Groves, Huang \& Chang]{groves1989geometric}
{\sc \au{Groves, N.~C.}, \au{Huang, T.~T.} \& \au{Chang, M.~S.}} \yr{1989}  \bt{Geometric characteristics of {DARPA} (defense advanced research projects agency) {SUBOFF} models ({DTRC} model numbers 5470 and 5471)}. Shd-1298-01.  \org{David Taylor Research Center}.

\bibitem[Jia {\em et~al.\/}(2022)Jia, Zou, Pang, Miao \& Li]{Jia2022Experimental}
{\sc \au{Jia, D.}, \au{Zou, Y.~C.}, \au{Pang, F.~Z.}, \au{Miao, X.~H.} \& \au{Li, H.~C.}} \yr{2022}  \at{Experimental study on the characteristics of flow-induced structure noise of underwater vehicle}.  \jt{Ocean Eng.}  \bvol{262},  \pg{112126}.

\bibitem[Jiang {\em et~al.\/}(2025{\natexlab{{\em a\/}}})Jiang, Liao, Liu \& Xie]{Jiang2025PoFSUBOFF}
{\sc \au{Jiang, P.}, \au{Liao, S.~J.}, \au{Liu, L.} \& \au{Xie, B.}} \yr{2025{\natexlab{{\em a\/}}}}  \at{Effects of appendages on the turbulence and flow noise of a submarine model using high-order scheme}.  \jt{Phys. Fluids}  \bvol{37},  \pg{095199}.

\bibitem[Jiang {\em et~al.\/}(2024)Jiang, Liao \& Xie]{Jiang2024OEHUll}
{\sc \au{Jiang, P.}, \au{Liao, S.~J.} \& \au{Xie, B.}} \yr{2024}  \at{Large-eddy simulation of flow noise from turbulent flows past an axisymmetric hull using high-order schemes}.  \jt{Ocean Eng.}  \bvol{312},  \pg{119150}.

\bibitem[Jiang {\em et~al.\/}(2025{\natexlab{{\em b\/}}})Jiang, Zhang, Dai, Peng, Xie \& Liao]{Jiang2025HullExperiment}
{\sc \au{Jiang, P.}, \au{Zhang, H.~Y.}, \au{Dai, Y.}, \au{Peng, T.}, \au{Xie, B.} \& \au{Liao, S.~J.}} \yr{2025{\natexlab{{\em b\/}}}}  \at{Experimental study on the wall-pressure fluctuations of flow over an axisymmetric hull}.  \jt{arXiv} Https://arxiv.org/abs/2512.22299,  \arxiv{arXiv: 2512.22299}.

\bibitem[Jim{\'e}nez {\em et~al.\/}(2010)Jim{\'e}nez, Hultmark \& Smits]{Jimenez2010JFMwakeflow}
{\sc \au{Jim{\'e}nez, J.~M.}, \au{Hultmark, M.} \& \au{Smits, A.~J.}} \yr{2010}  \at{The intermediate wake of a body of revolution at high reynolds numbers}.  \jt{J. Fluid Mech.}  \bvol{659},  \pg{516--539}.

\bibitem[Joseph(2017)]{Joseph2017Pressure}
{\sc \au{Joseph, L.~A.}} \yr{2017}  \at{Pressure fluctuations in a high-reynolds-number turbulent boundary layer over rough surfaces of different configurations}. PhD thesis, Virginia Polytechnic Institute and State University, Blacksburg, Virginia.

\bibitem[Joseph {\em et~al.\/}(2020)Joseph, Molinaro, Davenport \& Meyers]{Joseph2020JFMPlate}
{\sc \au{Joseph, L.~A.}, \au{Molinaro, N.~J.}, \au{Davenport, W.~J.} \& \au{Meyers, T.~W.}} \yr{2020}  \at{Characteristics of the pressure fluctuations generated in turbulent boundary layers over rough surfaces}.  \jt{J. Fluid Mech.}  \bvol{883},  \pg{A3}.

\bibitem[Kraichnan(1964)]{kraichnan1964kolmogorovs}
{\sc \au{Kraichnan, R.~H.}} \yr{1964}  \at{Kolmogorov's hypotheses and {Eulerian} turbulence theory}.  \jt{Phys. Fluids}  \bvol{7}~(11),  \pg{1723--1734}.

\bibitem[Leong {\em et~al.\/}(2016)Leong, Piccolini, Desjuzeur, Ranmuthugala \& Renilson]{leong2016evaluation}
{\sc \au{Leong, Zhi}, \au{Piccolini, Simon}, \au{Desjuzeur, Marc}, \au{Ranmuthugala, Dev} \& \au{Renilson, Martin}} \yr{2016} Evaluation of the out-of-plane loads on a submarine undergoing a steady turn.  \bt{In {\em 20th Australasian Fluid Mechanics Conference\/}}. Perth, Australia.

\bibitem[Liu {\em et~al.\/}(2023)Liu, Hao, Bie, Wang, Ren \& Hua]{liu2023control}
{\sc \au{Liu, Gang}, \au{Hao, Zongrui}, \au{Bie, Haiyan}, \au{Wang, Yue}, \au{Ren, Wanlong} \& \au{Hua, Zhili}} \yr{2023}  \at{Control mechanism of a vortex control baffle for the horseshoe vortex around the sail of a {DARPA} {SUBOFF} model}.  \jt{Ocean Eng.}  \bvol{275},  \pg{114166}.

\bibitem[Liu \& Huang(1998)]{Liu1998SUBOFFexp}
{\sc \au{Liu, H.} \& \au{Huang, T.}} \yr{1998}  \bt{Summary of {DARPA} {SUBOFF} experimental program data}. Report CRDKNSWC/HD-1298-11.  \org{Naval Surface Warfare Center, Carderock Division, Hydrodynamics Directorate}, Bethesda, MD.

\bibitem[Liu {\em et~al.\/}(2011)Liu, Xiong \& Tu]{liu2011numerical}
{\sc \au{Liu, Zhihua}, \au{Xiong, Ying} \& \au{Tu, Chengxu}} \yr{2011}  \at{Numerical simulation and control of horseshoe vortex around an appendage-body junction}.  \jt{J. Fluids Struct.}  \bvol{27},  \pg{23--42}.

\bibitem[Liu {\em et~al.\/}(2014)Liu, Xiong \& Tu]{Liu2014Method}
{\sc \au{Liu, Zhi~Hua}, \au{Xiong, Ying} \& \au{Tu, Cheng~Xu}} \yr{2014}  \at{The method to control the submarine horseshoe vortex by breaking the vortex core}.  \jt{J. Hydrodyn.}  \bvol{26}~(4),  \pg{637--645}.

\bibitem[Liu {\em et~al.\/}(2010)Liu, Xiong, Wang, Wang \& Tu]{Liu2010Numerical}
{\sc \au{Liu, Zhi~Hua}, \au{Xiong, Ying}, \au{Wang, Zhan~Zhi}, \au{Wang, Song} \& \au{Tu, Cheng~Xu}} \yr{2010}  \at{Numerical simulation and experimental study of the new method of horseshoe vortex control}.  \jt{J. Hydrodyn.}  \bvol{22}~(4),  \pg{572--581}.

\bibitem[Ma {\em et~al.\/}(2025)Ma, Li, Guo, Liao \& Yang]{ma2025performance}
{\sc \au{Ma, Zhi~Hao}, \au{Li, Peng}, \au{Guo, Hang}, \au{Liao, Kuai} \& \au{Yang, Yi~Ren}} \yr{2025}  \at{Performance and mechanism of the hydrodynamic noise reduction for biomimetic trailing-edge serrations of a submarine}.  \jt{J. Fluids Struct.}  \bvol{133},  \pg{104256}.

\bibitem[Meyers {\em et~al.\/}(2015)Meyers, Forest \& Davenport]{Meyers2015Wallpressure}
{\sc \au{Meyers, T.}, \au{Forest, J.~B.} \& \au{Davenport, W.~J.}} \yr{2015}  \at{The wall-pressure spectrum of {high-Reynolds-number} turbulent boundary-layer flows over rough surfaces}.  \jt{J. Fluid Mech.}  \bvol{768},  \pg{261--293}.

\bibitem[Morse \& Mahesh(2023)]{MorseMahesh2023Trippingeffects}
{\sc \au{Morse, N.} \& \au{Mahesh, K.}} \yr{2023}  \at{Tripping effects on model-scale studies of flow over the {DARPA} {SUBOFF}}.  \jt{J. Fluid Mech.}  \bvol{975}~(A3).

\bibitem[Posa \& Balaras(2016)]{posa2016suboff}
{\sc \au{Posa, A.} \& \au{Balaras, E.}} \yr{2016}  \at{\text{A} numerical investigation of the wake of an axisymmetric body with appendages}.  \jt{J.~Fluid~Mech.}  \bvol{792},  \pg{470--498}.

\bibitem[Rahmani {\em et~al.\/}(2023)Rahmani, Gavzan \& Manshadi]{rahmani2023experimental}
{\sc \au{Rahmani, Mohsen}, \au{Gavzan, Iraj~Jafari} \& \au{Manshadi, Mojtaba~Dehghan}} \yr{2023}  \at{Experimental investigation of the effect of sail geometry on the flow around the suboff submarine model inspired by the dolphin's dorsal fin}.  \jt{Ships Offshore Struct.} .

\bibitem[Toxoepus {\em et~al.\/}(2014)Toxoepus, Kuin, Kerkvliet, Hoeijmakers \& Nienhuis]{toxopeus2014improvement}
{\sc \au{Toxoepus, Serge}, \au{Kuin, Roderik}, \au{Kerkvliet, Maarten}, \au{Hoeijmakers, Harry} \& \au{Nienhuis, Bart}} \yr{2014} Improvement of resistance and wake field of an underwater vehicle by optimising the fin-body junction flow with {CFD}.  \bt{In {\em Proceedings of the ASME 2014 33rd International Conference on Ocean, Offshore and Arctic Engineering (OMAE2014)\/}}. San Francisco, California, USA.

\bibitem[Tsuji {\em et~al.\/}(2012)Tsuji, Imayama, Schlatter, Alfredsson, Johansson, Marusic, Hutchins \& Monty]{Tsuji2012Pressure}
{\sc \au{Tsuji, Y.}, \au{Imayama, S.}, \au{Schlatter, P.}, \au{Alfredsson, P.~H.}, \au{Johansson, A.~V.}, \au{Marusic, I.}, \au{Hutchins, N.} \& \au{Monty, J.}} \yr{2012}  \at{Pressure fluctuation in {high-Reynolds-number} turbulent boundary layer: {Results} from experiments and {DNS}}.  \jt{J. Turbul.}  \bvol{13},  \pg{N50}.

\bibitem[Wang {\em et~al.\/}(2021)Wang, Huang \& Pan]{wang2021numerical}
{\sc \au{Wang, Xi~Hui}, \au{Huang, Qiao~Gao} \& \au{Pan, Guang}} \yr{2021}  \at{Numerical research on the influence of sail leading edge shapes on the hydrodynamic noise of a submarine}.  \jt{Appl. Ocean Res.}  \bvol{117},  \pg{102935}.

\bibitem[Willmarth(1975)]{Willmarth1975Pressure}
{\sc \au{Willmarth, W.~W.}} \yr{1975}  \at{Pressure fluctuations beneath turbulent boundary layers}.  \jt{Annu. Rev. Fluid Mech.}  \bvol{7},  \pg{13--38}.

\bibitem[Wu {\em et~al.\/}(2024)Wu, Zhang, Sun \& Xiao]{Wu2023VCBCFD}
{\sc \au{Wu, Guibin}, \au{Zhang, Guiyong}, \au{Sun, Tiezhi} \& \au{Xiao, Qihang}} \yr{2024}  \at{Study on optimization of submarine wake field based on vortex control baffle}.  \jt{J. Huazhong Univ. Sci. Tech. (Nat. Sci. Ed.)} (In Chinese).

\bibitem[Yang \& Yang(2022)]{Yang_Yang_2022}
{\sc \au{Yang, B.~W.} \& \au{Yang, Z.~X.}} \yr{2022}  \at{On the wavenumber-frequency spectrum of the wall pressure fluctuations in turbulent channel flow}.  \jt{J. Fluid Mech.}  \bvol{937},  \pg{A39}.

\bibitem[Yu {\em et~al.\/}(2007)Yu, Wu \& Pang]{Yu2007shipNoise}
{\sc \au{Yu, M.~S.}, \au{Wu, Y.~S.} \& \au{Pang, Y.~Z.}} \yr{2007}  \at{A review of progress for hydrodynamic noise of ships}.  \jt{J. Ship Mech.}  \bvol{11}~(1),  \pg{152--158}, (in Chinese).

\bibitem[Zhang \& Xu(2020)]{zhang2020review}
{\sc \au{Zhang, W.~W.} \& \au{Xu, R.~W.}} \yr{2020}  \at{Review of research on sail hydrodynamic noise and control technology}.  \jt{Chin. J. Ship Res.}  \bvol{15}~(6),  \pg{72--89}.

\bibitem[Zhou {\em et~al.\/}(2022)Zhou, Xu, Wang \& He]{Zhou2022suboff}
{\sc \au{Zhou, Z.~T.}, \au{Xu, Z.~Y.}, \au{Wang, S.~Z.} \& \au{He, G.~W.}} \yr{2022}  \at{Wall-modeled large-eddy simulation of noise generated by turbulence around an appended axisymmetric body of revolution}.  \jt{J. Hydrodyn.}  \bvol{34}~(4),  \pg{533--554}.

\end{thebibliography}

\end{document}